\PassOptionsToPackage{dvipsnames}{xcolor}
\documentclass[twocolumn,twcolappendix,times]{aastex631}

\usepackage{soul}

\newcommand{\revision}[1]{#1}

\newcommand{\vect}[1]{\boldsymbol{#1}} 

\shorttitle{SPIn4D}
\shortauthors{Yang et al.}

\usepackage{amsmath}
\usepackage{graphicx}
\usepackage{CJK}

\makeatletter
\newcommand*{\rom}[1]{\expandafter\@slowromancap\romannumeral #1@}
\makeatother

\graphicspath{{./}{/}}

\begin{document}
\begin{CJK*}{UTF8}{gbsn}

\title{Spectropolarimetric Inversion in Four Dimensions with Deep Learning (SPIn4D): \\
	\rom{1}. Overview, Magnetohydrodynamic Modeling, and Stokes Profile Synthesis}

\author[0000-0002-7663-7652]{Kai E. Yang (杨凯)}
\affiliation{Institute for Astronomy, University of Hawai`i at M\=anoa, Pukalani, HI 96768, USA}

\author[0000-0002-8259-8303]{Lucas A. Tarr}
\affiliation{National Solar Observatory, Pukalani, HI 96768, USA}

\author[0000-0001-5850-3119]{Matthias Rempel}
\affiliation{High Altitude Observatory, NSF National Center for Atmospheric Research, Boulder, CO 80307, USA}

\author[0000-0001-6311-146X]{S. Curt Dodds}
\affiliation{Institute for Astronomy, University of Hawai`i at M\=anoa, Honolulu, HI 96822, USA}

\author[0000-0001-5459-2628]{Sarah A. Jaeggli}
\affiliation{National Solar Observatory, Pukalani, HI 96768, USA}

\author[0000-0002-7354-5461]{Peter Sadowski}
\affiliation{Department of Information and Computer Sciences, University of Hawai`i at M\=anoa, Honolulu, HI 96822, USA}

\author[0000-0002-7451-9804]{Thomas A. Schad}
\affiliation{National Solar Observatory, Pukalani, HI 96768, USA}

\author[0000-0001-5171-9144]{Ian Cunnyngham}
\affiliation{Institute for Astronomy, University of Hawai`i at M\=anoa, Honolulu, HI 96822, USA}

\author[0000-0002-7290-0863]{Jiayi Liu (刘嘉奕)}
\affiliation{Institute for Astronomy, University of Hawai`i at M\=anoa, Honolulu, HI 96822, USA}

\author[0000-0001-7217-9749]{Yannik Glaser}
\affiliation{Department of Information and Computer Sciences, University of Hawai`i at M\=anoa, Honolulu, HI 96822, USA}

\author[0000-0003-4043-616X]{Xudong Sun (孙旭东)}
\affiliation{Institute for Astronomy, University of Hawai`i at M\=anoa, Pukalani, HI 96768, USA}

\correspondingauthor{Xudong Sun (孙旭东)}
\email{xudongs@hawaii.edu}


\begin{abstract}

The National Science Foundation's \textit{Daniel K. Inouye Solar Telescope} (\textit{DKIST}) will provide high-resolution, multi-line spectropolarimetric observations that are poised to revolutionize our understanding of the Sun. Given the massive data volume, novel inference techniques are required to unlock its full potential. Here, we provide an overview of our ``SPIn4D'' project, which aims to develop deep convolutional neural networks (CNNs) for estimating the physical properties of the solar photosphere from \textit{DKIST} spectropolarimetric observations. We describe the magnetohydrodynamic (MHD) modeling and the Stokes profile synthesis pipeline that produce the simulated output and input data, respectively. These data will be used to train a set of CNNs that can rapidly infer the four-dimensional MHD state vectors by exploiting the spatiotemporally coherent patterns in the Stokes profile time series. Specifically, our radiative MHD model simulates the small-scale dynamo actions that are prevalent in quiet-Sun and plage regions. Six cases with different mean magnetic fields have been conducted; each case covers six solar-hours, totaling $109$~TB
in data volume. The simulation domain covers at least $25\times25\times8$~Mm with $16\times16\times12$~km spatial resolution, extending from the upper convection zone up to the temperature minimum region. The outputs are stored at a 40~s cadence. We forward model the Stokes profile of two sets of \ion{Fe}{1} lines at $630$ and $1565$~nm, which will be simultaneously observed by \textit{DKIST} and can better constrain the parameter variations along the line of sight. The MHD model output and the synthetic Stokes profiles are publicly available, \revision{with 13.7~TB in the initial release}.

\end{abstract}



\section{Introduction} \label{sec:intro}
The solar photosphere, a dynamic layer characterized by diverse plasma and magnetic states, plays an essential role in solar astronomy. 
Within this layer, processes such as magnetic field emergence, convection, and energy and helicity injection are continuously ongoing and define the structure and evolution of the outer solar atmosphere and the rest of the heliosphere.
Understanding the photospheric plasma is therefore key to understanding solar activity.
The photosphere is dynamically described by the magnetohydrodynamic \citep[MHD,][]{Pries2014book} equations, which govern the four-dimensional (4D; three for space plus one for time) evolution of the MHD state vector: the magnetic $\vect{B}$ and velocity $\vect{v}$ vector fields, and the scalar density $\rho$ and pressure $p$ (or temperature $T$) fields.
In addition to the MHD equations, radiative transfer plays a dual role in the system.  On the one hand, it directly modifies the thermodynamic evolution of the plasma relative to basic MHD via the emission and absorption of photons, while on the other hand, the resulting spectra that reaches the far field (e.g., at ground and space-based telescopes) encodes the state of the system, and thus provides crucial diagnostics of the photospheric plasma \citep{Hubeny2015book}.
Notably, the Zeeman and Hanle effects encode the state of magnetic field in \emph{polarized} radiation, provided it can be properly interpreted \citep{Stenflo1994ASSL,delToroIniesta2007book}.

The \textit{Daniel K. Inouye Solar Telescope} \citep[\textit{DKIST},][]{Rimmele2020SoPh}, established by the National Science Foundation, has opened a new era in solar observation with its commencement of scientific operations in 2022.
Boasting the world's largest aperture for solar studies (4~m), \textit{DKIST} enables unprecedentedly sharp observations (down to 0.03\arcsec, or 25~km) with the excellent seeing conditions on Haleakal\=a and its advanced adaptive optics system.  
Both the design of the telescope and its instrumentation suite enable measurements of polarized signals with unparalleled accuracy \citep[approximately $5\times10^{-4}$,][]{Rimmele2020SoPh,deWijn2022SoPh,Jaeggli2022SoPh,Harrington2023SoPh}. 
Its large aperture means it can match the signal-to-noise ratio (SNR) of deep-exposure quiet Sun observations by the \textit{Hinode}/Spectro-Polarimeter \citep{lites2008} in a fraction of the time ($\sim$$60$ s versus $\sim$$1$ s). 
Equipped with cutting-edge instruments like the Visible Spectro-Polarimeter \citep[ViSP,][]{deWijn2022SoPh}, the Diffraction-Limited Near-Infrared Spectro-Polarimeter \citep[DL-NIRSP,][]{Jaeggli2022SoPh}, the Cryogenic Near-Infrared Spectropolarimeter \citep[Cryo-NIRSP,][]{Schad2023ApJ}, and the Visible Tunable Filter \citep[VTF,][]{Schmidt2016SPIE}, \textit{DKIST} has the capability to explore various solar regions, from the photosphere and chromosphere to the corona. 
Here we focus on the capabilities of the DL-NIRSP, which currently observes in the near-infrared spectrum using integrating fiber-optic integral-field units (IFU) that enable the simultaneous collection of multi-line Stokes profiles across a continuous field of view (FOV)\footnote{DL-NIRSP recently (winter 2023) swapped the fiber-optic IFU for a newly developed image-slicer.  The resulting data products will be functionally equivalent for our purposes, but have increased fidelity.}. 
This feature significantly enhances our ability to measure and analyze solar phenomena in detail, particularly the magnetic properties of the small-scale structures.

With the help of the above new facilities, spectropolarimetric observations (typically in the form of wavelength dependent Stokes profiles) for solar physics will significantly expand in the near future. These advanced observations are sensitive to the plasma state variables across the solar atmosphere \citep[][and reference therein]{delToroIniesta2016,BellotRubio2019LRSP}, offering a unique tool to probe the solar environment in different heights through sophisticated inversion techniques, whereby the multi-dimensional state of the plasma is inferred from 2D maps of the polarized spectra \citep{RuizCobo1992,AsensioRamos2008ApJ,SocasNavarro2015AA,Milic2018AA,delaCruzRodriguez2019AA,QuinteroNoda2021AA,QuinteroNoda2023AA,RuizCobo2022,asensioramos2019,PastorYabar2019AA,Borrero2019AA}. 
For example, initial \textit{DKIST}/ViSP data measurements of magnetic fields in both quiet and plage regions highlight the utility of spectropolarimetric observation in probing plasma dynamics in otherwise inaccessible environments \citep{Campbell2023ApJ,daSilvaSantos2023ApJ,Kuridze2024}.  Developing good inversion techniques is thus required to tackle a host of unanswered questions in both solar and plasma physics in general, for instance, probing the mechanisms behind local magnetic dynamos \citep{Vogler2007AA,Stenflo2012AA,rempel2014,Lord2014ApJ}, understanding the flux of helicity and energy across the solar surface \citep{welsch2015,schuck2019,Lumme2019SoPh,Thalmann2021ApJ,Liu2012ApJ,Liu2014ApJ,Liu2023ApJ}, and unraveling the magnetic foundations necessary for solar eruptions and coronal heating \citep{Antiochos1998ApJ,Antiochos1999ApJ,Moore2001ApJ,Priest2002AARv,Sun2013ApJ,Wang2015NatCo,Liu2016NatCo,Chitta2017AA,Wyper2017Natur,Wang2017NatAs,Samanta2019Sci}. 

\begin{figure*}
\centering
\fig{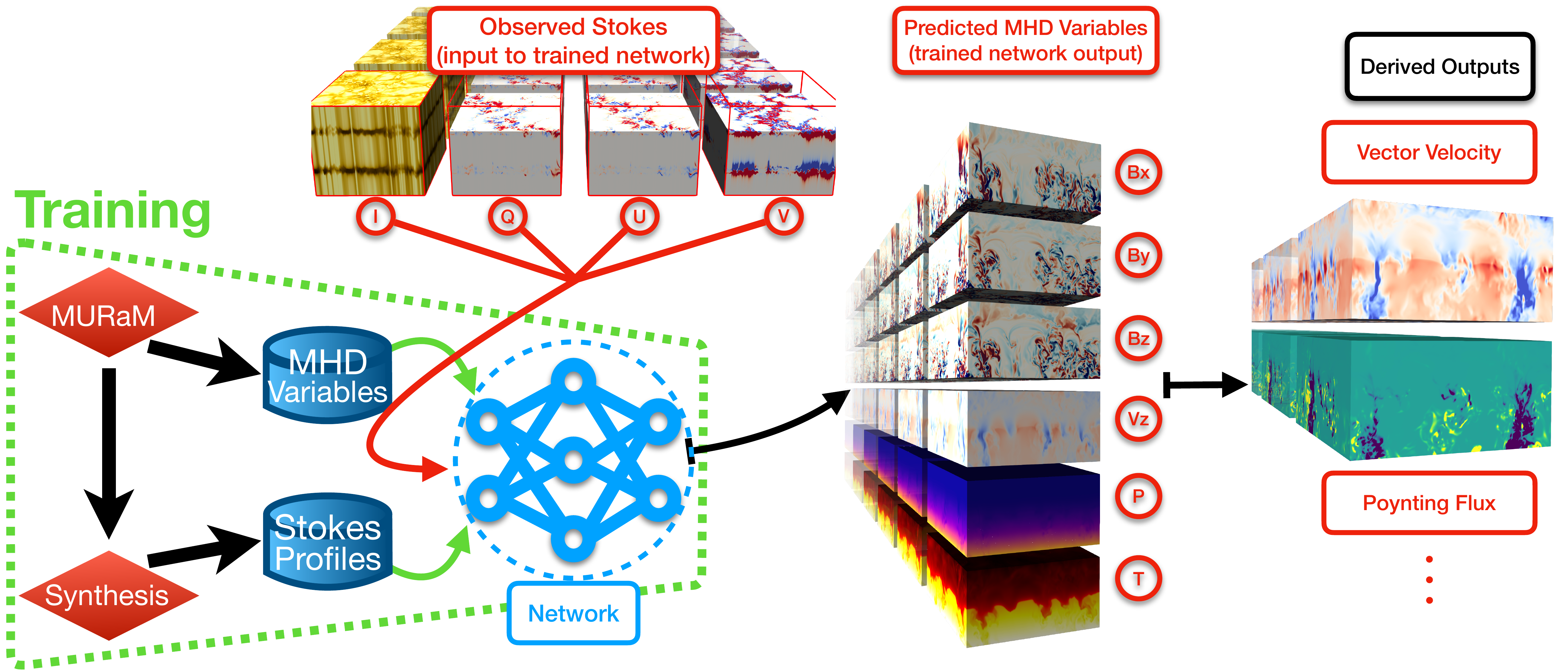}{\textwidth}{}
\caption{Schematic representation of the SPIn4D model workflow. The core of the model is the DL neural network, highlighted in blue in the middle of the diagram.  The network Training step is outlined by the broken green line and uses data derived from the MURaM simulations, both the MHD variables themselves and the Stokes profiles $(I,Q,U,V)$ synthesized from the MHD data cubes (green lines). 
Once trained on the simulated data, Observed Stokes data can be input to the network (red arrow) to produce the most likely 3D MHD state as output (labeled "Predicted MHD Variables").
The network may be trained to receive single-time input Stokes data to produce a reduced dimensional output MHD state $(\vect{B}, v_z, P, T)$ or to receive multi-time input Stokes data to produce a full-dimensional MHD output, 
including additional Derived Outputs such as vector velocities, Poynting flux, and so on. The network may also be trained to produce the Derived Outputs directly.\label{fig:0}}
\end{figure*}

Despite recent advancements in the inversion technology, creating accurate 3D reconstructions of the solar atmosphere still remains a challenge. In particular, knowledge of the physical variables' distribution on a spatial grid is crucial for the calculation of differential quantities that define key parameters of the system, e.g., the electric current, Lorentz force, helicity and energy flux, and so on. Most current inversion techniques return the physical state of the plasma on an optical depth $\tau$ grid rather than on a spatial grid. Converting to a physical grid along the line of sight (LOS) involves the additional steps of resolving the ambiguity in the inverted magnetic azimuth angle, reconstructing the atmosphere based on dynamic/static assumptions, and defining a vertical offset between each (assumed) independent LOS \citep{PastorYabar2019AA,Borrero2019AA,Borrero2021AA,Borrero2023AA}.
The spatiotemporal information of the time-series observations is governed by the MHD equations. Once incorporated, they can help resolve the LOS spatial grid and ambiguity issues while also enhancing the physical accuracy of inversions. However, integrating this information is challenging in traditional 1D inversion techniques based on sophisticated radiative transfer methods.
These challenges are exacerbated by the need for powerful computers to process the data and the complex work of creating accurate solar models that match up with scientific laws over time, e.g., Newton's laws and Maxwell's equations.  Moreover, the arrival of powerful telescopes like \textit{DKIST}, which can generate about 20~TB of data every day, makes these challenges even greater, testing the limits of our current technology. This huge amount of data highlights the urgent need for innovative, efficient methods to handle and analyze large-volume solar Stokes profile data in the \textit{DKIST} era.

On the other hand, deep learning (DL), a specialized branch of machine learning (ML), has shown exceptional efficacy in deriving approximate inferences from physics models \citep{Sadowski2018,Brehmer2020PNAS}. The rapid progress of ML in solar physics, especially through DL's application to Stokes inversion, would significantly enhance our understanding and analytical abilities \citep[][and reference therein]{AsensioRamos2023LRSP}. Recent studies have highlighted the effectiveness of convolutional neural networks \citep[CNNs,][]{asensioramos2019,Milic2020AA,Gafeira2021AA,Centeno2022ApJ,Chappell2022AA,Higgins2021ApJ,Higgins2022ApJS,Mistryukova2023SoPh,Rahman2023ApJ,Rahman2024ApJS} and other advanced ML algorithms \citep{bobra2015,bobra2016,Florios2018SoPh,Huang2018ApJ,Nishizuka2018ApJ,SainzDalda2019ApJ,DiazBaso2022AA,VicenteArevalo2022ApJ,Jarolim2023NatAs,Jarolim2024ApJa,Jarolim2024ApJb,Goodwin2024arXiv} in processing a wide array of solar observations, ranging from the quiet Sun to dynamic solar flares and coronal mass ejections. These models have shown promising results in enhancing accuracy and efficiency, notably outpacing traditional methods in speed without sacrificing analytical complexity.
Furthermore, the DL methods can enhance spatial resolution and image denoising, improving the capability of observing small structures \citep{DiazBaso2018A&A...614A...5D,DiazBaso2019A&A...629A..99D,AsensioRamos2018A&A...620A..73A,Rahman2020ApJ...897L..32R,Song2022ApJS..263...25S,Eklund2023A&A...669A.106E}. With the help of time-series observations, it also effectively resolves solar surface flows, with the DeepVel code \citep{asensioramos2017} demonstrating superior performance compared to conventional methods in analyzing these small-scale structures \citep{Tremblay2018SoPh..293...57T}.

Most of the aforementioned ML models are supervised learning models, requiring a large volume of training data to derive a relationship between the input and target data.  Fortunately, modern numerical MHD simulations can now accurately mimic various solar phenomena \citep{rempel2012,rempel2014,cheung2010,cheung2019,ChenF2017,ChenF2023ApJ,ChenF2023ApJL} and therefore allow the generation of extensive and realistic datasets for DL training. Once trained, DL inversion models are extremely fast to run. For example, in the pioneering work of \cite{asensioramos2019}, they train 2D CNN models using a radiation-MHD simulation of a sunspot and synthesized Stokes profiles.  When applied to observational data from the \textit{Hinode}/SP, their CNN model inverts a $512\times512$ pixel map within $\sim$$180$ ms, orders of magnitude faster than current inversion methods.  Additionally, the model could recover the 3D MHD variables on a true spatial grid rather than an optical depth grid, and at roughly half the error compared to the Stokes Inversion based on Response functions \citep[SIR,][]{RuizCobo1992,ruizcobo2012} code. It is worth mentioning that \citet{asensioramos2019} did not treat the azimuthal ambiguity in the magnetic field directly, but instead solved a reduced problem using a coordinate transformation.

To address the challenges outlined above and take advantage of the potential of DL, we launched the ``Spectropolarimetric Inversion in Four Dimensions with Deep Learning'' (SPIn4D) project\footnote{\url{https://ifauh.github.io/SPIN4D/}}. to train CNN models on radiative MHD (RMHD) simulations with a significantly larger dataset, \revision{i.e., larger domain size, longer evolution, and more cases with a variety of mean magnetic field strengths,}
compared to previous efforts. \revision{As discussed below, this provides more statistically independent snapshots useful for training.} Detailed information about the data, including access methods, are publicly available\footnote{\revision{Several data access methods are offered for Data Release 1 at \url{http://dtn-itc.ifa.hawaii.edu/spin4d/DR1/}.}}.

\autoref{fig:0} presents a schematic overview of key elements of each phase of the project.  Our focus will be on photospheric regions with intermediate field strengths (between quiet sun and plage, or up to a few hundred Gauss when spatially averaged), characterized by relatively simple field geometries. These regions are expected to be prevalent in the initial years of \textit{DKIST}'s operation. Simulations are carried out using the Max-Planck University-of-Chicago Radiative MHD code \citep[MURaM,][]{Vogler2005AA,Rempel2009ApJ,cheung2010,rempel2012,rempel2014} and take the quiet Sun small-scale dynamo simulation (SSD) of \citet{rempel2014} as the point of departure. \revision{The fact that these MURaM simulations use gray radiative transfer is not expected to impact the results here. As both the training and the evaluation steps use synthetic data, the ML models should simply learn the mapping between the self-consistent inputs and outputs, without regards to the detailed physics. When applying to real observations, however, we do expect ML models trained on more realistic, non-gray radiative transfer simulations \citep{Vogler2004A&A...421..755V} to perform better, at the expense of increased computational costs.}

Next, we 
employ forward-modeling of radiative transfer through these simulations using a new version of the SIR code (SIR3D) and the Departure Coefficient aided Stokes Inversion based on Response Functions \citep[DeSIRe,][]{RuizCobo2022} code to synthesize Stokes profiles for multi-line observations. We have selected the well-studied \ion{Fe}{1} lines at $630.15$ nm, $630.25$ nm, $1564.9$ nm, and $1565.2$~nm due to their significant Land\'e factors ($2.5$, $1.67$, $3$, and $1.53$, respectively) and their wide-spread use in ground and space-based observations with high Stokes SNR. 

The third phase involves developing CNN models that aim to accurately correlate time series of observational data, especially those from the DL-NIRSP instrument at \textit{DKIST}, with precise 4D MHD states. The models will be rigorously trained and evaluated using the data generated in the first two steps. The ability of this model will be compared with the SIR inversion code as a base line. 
Higher-level variables such the vector velocity and associated Poynting flux across the photosphere may \revision{be estimated either based on the inversion results as in conventional methods \citep[e.g.,][]{liuy2012,kazachenko2014}, or ultimately directly from our ML models that encode the temporal information with a time-series of Stokes profiles as input}.

In this paper, we focus on the generation of the training data for the SPIn4D project.  The paper is organized as follows: Section~\ref{sec:2} delves into the specifics of the MURaM SSD simulation. The process of multi-line synthesis is explored in Section~\ref{sec:3}. Section~\ref{sec:simulationproblems} addresses simulation artifacts and their treatment.
Finally, a summary is provided in Section~\ref{sec:4}. Additional details on the pipeline are included in the Appendix.

\begin{deluxetable}{cccc}	
\tablenum{1}
\tablecaption{MHD simulation summary\label{tab:1}}
\tablewidth{0pt}
\tablehead{
\colhead{ } & \colhead{Modified Field (G)\tablenotemark{a}}  & \colhead{Duration (hrs)\tablenotemark{b}}  & \colhead{Size (TB)\tablenotemark{c}}}
\startdata
Case 1  & 0 & 5.97 & 12\\
Case 2  &  $B_{Z}:50$& 6.21 &12\\
Case 3  &  $B_{X,Y,Z}:50$\tablenotemark{d}  &6.23 &12\\
Case 4  &  $B_{Z}:100$ & 6.02 & 12\\
Case 5  &  $B_{Z}:200$ & 6.04 & 12\\
Case 6  &  $B_{Z}:200,-150,-50$\tablenotemark{e} &6.20 &49
\enddata
\tablecomments{
\tablenotetext{a}{The magnetic field added to the relaxed SSD O16bM atmosphere as the initial condition for our simulations.}
\tablenotetext{b}{The physical time of the total simulations for each case.}
\tablenotetext{c}{The total size of the output 3D atmosphere files.}
\tablenotetext{d}{Case 3 has $B_X=B_Y=B_Z =+50$G.}
\tablenotetext{e}{Case 6 has $+200$ G, $-150$ G, and $-50$ G added to $B_Z$ in three quadrants (see \autoref{fig:2}).}
}
\end{deluxetable}

\section{Solar Atmosphere Simulation} \label{sec:2}
We ran six RMHD simulations covering a variety of photospheric conditions ranging from very quiet Sun to fairly strong plage.
Synthetic spectropolarimetric observations are created by running radiative forward models to through the RMHD output (Section~\ref{sec:3}).
The RMHD simulations use the MURaM code \citep{Vogler2005AA,Rempel2009ApJ} and take the relaxed solar atmosphere of the small scale dynamo case SSD 016bM from \citet{rempel2014} as their starting point.
Compared to \citet{rempel2014}, the simulation domain was extended by 500 km in the vertical direction above the photosphere and the vertical grid spacing was reduced from $16$ to $12$~km.
This formed the basis of our Case 1, a straightforward continuation of the SSD O16bM simulation. 
Cases 2, 4, and 5 simulate regions with increasingly stronger average field strength, introduced as an additional uniform magnetic field to each initial condition's vertical component, $B_Z$, at strengths of $50$ G, $100$ G, and $200$ G, respectively. 
Case 3 was augmented with an inclined initial magnetic field instead, uniformly set at $50$ G across all magnetic components.

The computational domain for Cases 1--5 spanned $24.6\times24.6\times8$ Mm, with a spatial resolution of $16\times16\times12$ km. The horizontal grid size is approximately half of the diffraction limit of \textit{DKIST}.
We saved data output every $40$ seconds to match the expected cadence from DL-NIRSP observations. 
The side boundaries of the domain are periodic, the bottom boundary is open for convective flows as detailed in \cite{rempel2014}, and the top boundary applies a potential magnetic field extrapolation in the ghost layers, along with a semi-transparent boundary condition for hydrodynamic variables, i.e.~density, velocity, and internal energy. This condition is designed to be open for outflows and closed for inflows, featuring (anti-)symmetric settings in the ghost layers for inflows and outflows, respectively.
Radiative transfer calculations may be carried out in arbitrary directions through the simulated domain, but for the tasks described in this work, we take the LOS direction to be the simulation $Z$ direction and use these two interchangeably.

\begin{figure*}
\centering
\fig{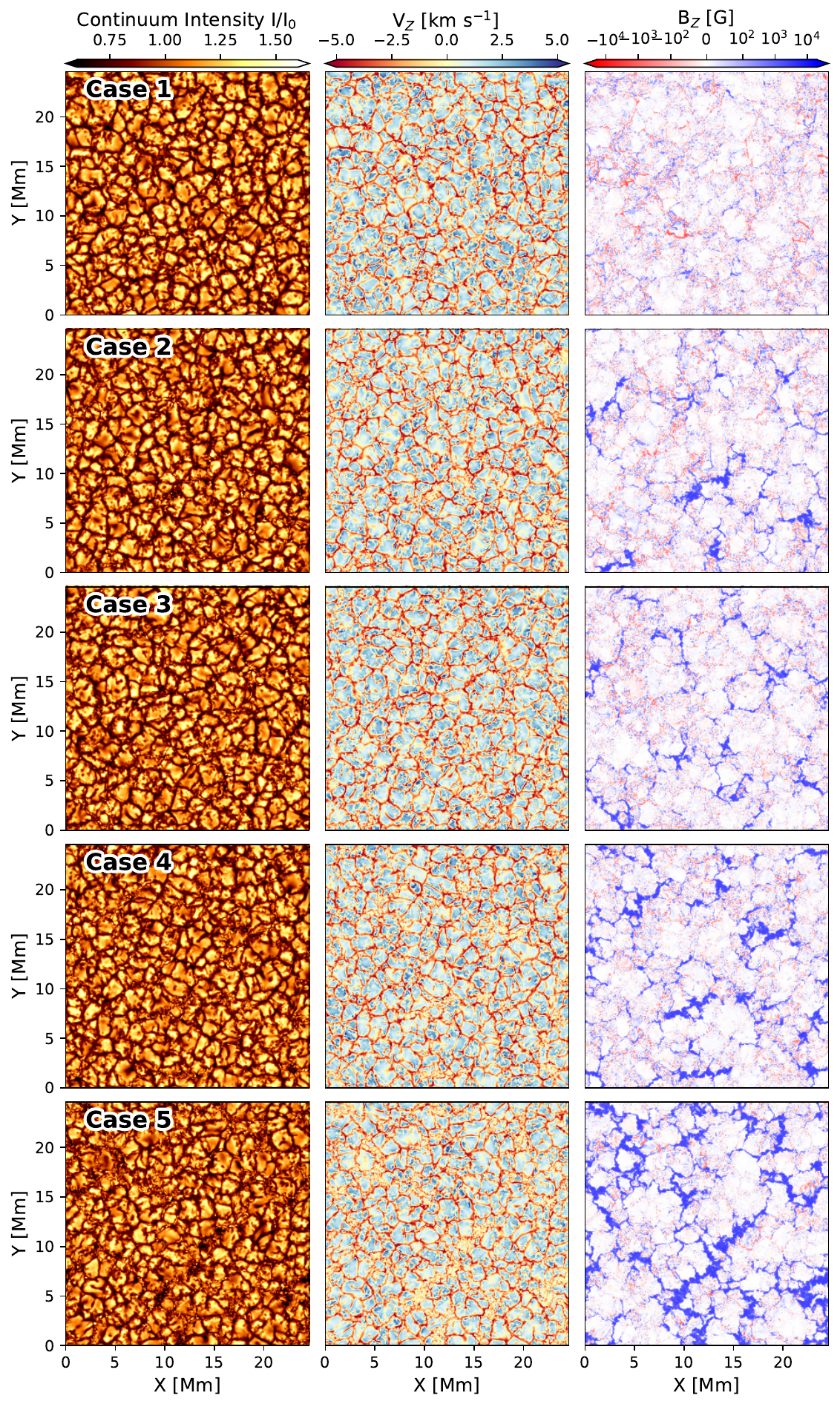}{0.7\textwidth}{}
\caption{From top to bottom, representative simulation results for Cases 1--5 at simulation times $5.43$ hrs, $5.59$ hrs, $5.59$ hrs, $5.59$ hrs, and $5.58$ hrs, respectively.  From left to right, the columns show continuum emission at 500~nm, LOS velocity $(V_Z)$, and LOS magnetic induction ($B_Z$) extracted surface of $\log_{10}\tau=0$, respectively. \label{fig:1}}
\end{figure*}

In \autoref{fig:1}, we show representative simulation output toward the end of each run for Cases 1--5: the $500$ nm continuum intensity (left), LOS velocity ($V_Z$; center), and LOS magnetic field ($B_Z$; right), each extracted from the layer of optical depth $\log_{10}\tau=0$ at roughly 5.5 hrs into each run, where $\tau$ is the optical depth at 500 nm.
The increasing initial average field strength in each case forms more and stronger field concentrations, as seen in the right column.
The greatest addition in Case 5 shows the development of several pores with substantially reduced intensity (left column) associated with the strongest field concentrations (right column), consistent with regions of strong plage on the Sun.

\begin{figure*}
\centering
\fig{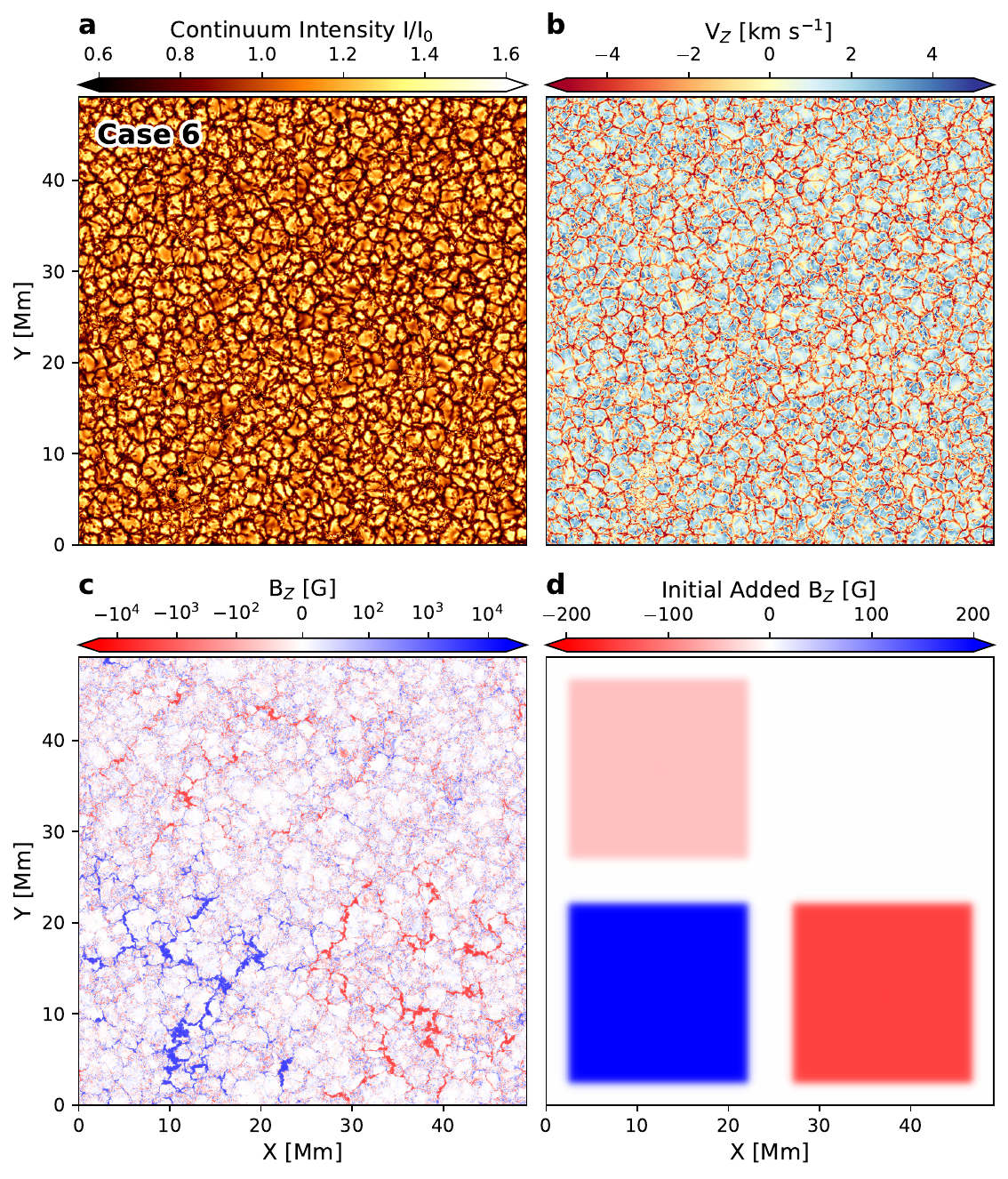}{\textwidth}{}
\caption{Similar to \autoref{fig:1}, but for Case 6 at $5.44$ hrs. Panel (d) shows the additional vertical field added to each quadrant and outlines the mask function.\label{fig:2}}
\end{figure*}

Cases 2--5 each have a net magnetic flux that would mimic coronal hole environments on the actual Sun, with no strong polarities with opposing signs. 
To account for real-Sun conditions with mixed polarities, we ran an additional Case 6 in a larger domain created by stitching together four of our modified SSD O16bM initial conditions in the two periodic horizontal directions and then adding additional field of strengths $200$ G, $-150$ G, and $-50$ G to the vertical component in three quadrants; the forth quadrant was left in the initial SSD configuration. 
To avoid discontinuous field strengths, the vertical field added to each quadrant was first multiplied by a mask function that decreases smoothly to zero at each quadrant boundary, 
\begin{align}
    \mathrm{mask}(x,y)&=\frac{1}{4}\biggl(\tanh\biggl[\frac{x-0.1}{0.02}\biggr]-\tanh\biggl[\frac{x-0.9}{0.02}\biggr]\biggr)\notag\\
    &\times \biggl(\tanh\biggl[\frac{y-0.1}{0.02}\biggr]-\tanh\biggl[\frac{y-0.9}{0.02}\biggr]\Biggr),
\end{align} 
where $x$ and $y$ are normalized coordinates in each field region before being stitched into the larger domain. 

Representative results for Case 6 are shown in \autoref{fig:2} around 5.5 hours into the simulation.
\autoref{fig:2}(d) shows the additional vertical field added to each quadrant in the initial condition.
After a few solar hours of relaxation, this simulation showed a mix of strong polarities with interaction of opposite polarity patches at the quadrant boundaries (see \autoref{fig:2}(c)).
Portions of this simulation may therefore represent the boundaries of long--decayed active regions.

\begin{figure*}[t!]
\centering
\fig{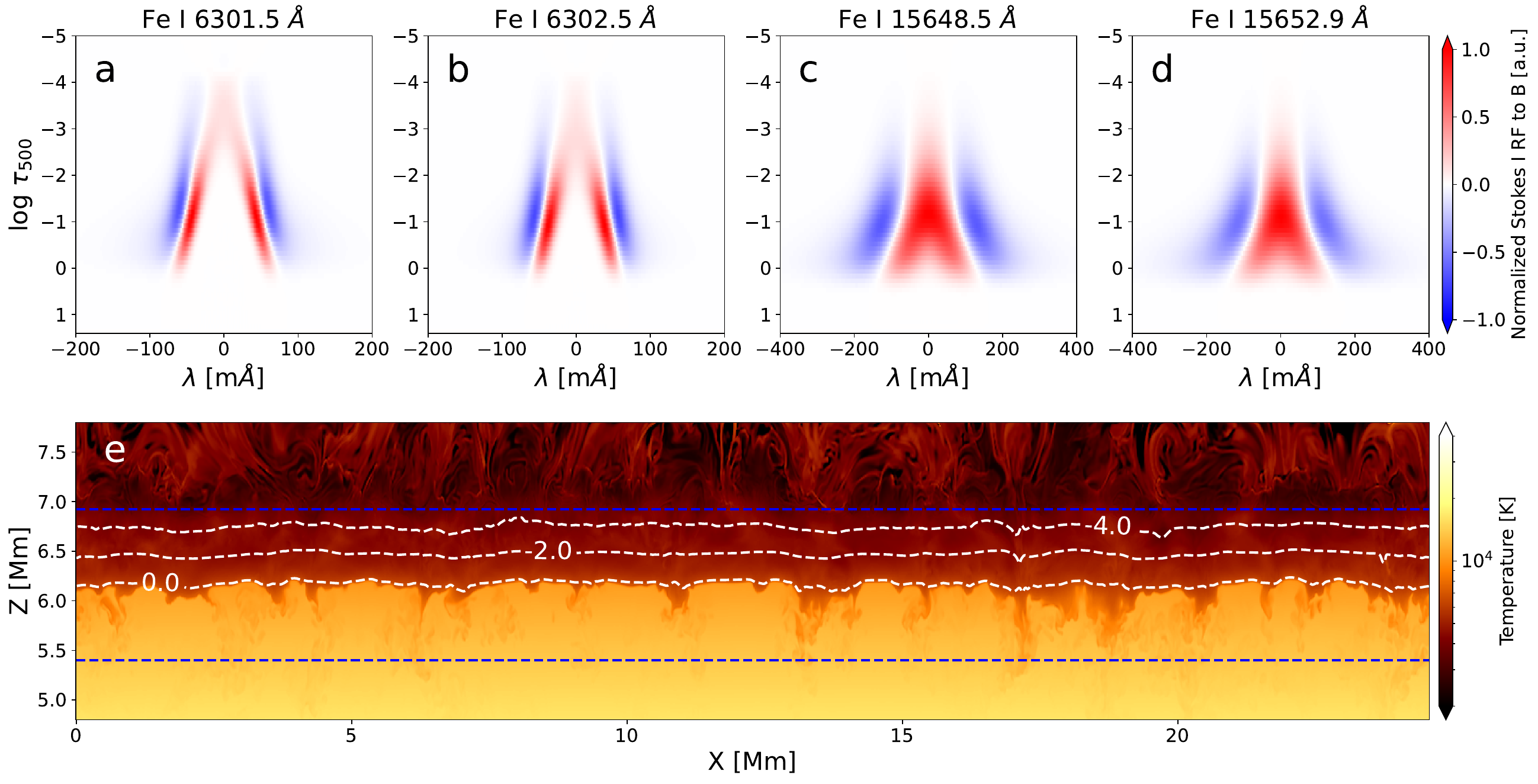}{\textwidth}{}
\caption{(a)--(d) The response function of Stokes $I$ to the magnetic induction for the four lines used in our study. (e) Temperature distribution on a vertical slice at the middle of the computational domain, $y=12.28$ Mm, from Case 1 at $t=5.43$ hrs. The three white dashed lines indicate the contours of the logarithm optical depth, $\log_{10}\tau$, with the values of $0$, $-2$, and $-4$. The two blue dashed lines indicate the sub-domain we extracted for the training dataset.
\label{fig:3}}
\end{figure*}

Each simulation extends over $8$ Mm vertically for all Cases, from a shallow convection zone to the upper photosphere, represented by $672$ grid points. 
The photospheric \ion{Fe}{1} lines that we are interested in form in a relatively narrow layer ($\sim 1$ Mm)  entirely contained within the simulation domain.
In \autoref{fig:3}(a)--(d) we show the response functions of Stokes $I$ to the magnetic field magnitudes of these lines calculated using SIR.
Based on this, and to optimize storage, we selected a range of $128$ layers from the simulation interiors for 
output (at 40~s cadence), specifically from the $z$ grid points $450$ to $577$ in the vertical direction, spanning an optical depth range from $10^5$ to $10^{-5}$ and demarcated by the blue dashed lines in \autoref{fig:3}(e). 
Comparison with the response functions shows that the extracted region adequately encompasses the formation layers of the \ion{Fe}{1} lines of interest. 
Further details are provided in Section \ref{sec:3}.

All Cases 1--6 were simulated for about six physical hours \revision{(including the initial two hours of relaxation)} and generated a total of $109$ TB of output for the 3D MHD variables, including the high-cadence output just described plus occasional snapshots of the full numerical domain. 
The initial magnetic configuration, total duration, and total output for all cases are summarized in \autoref{tab:1}.

\subsection{Initial relaxation, convective turnover time, and statistically independent training data}\label{sec:statisticalindependence}

\begin{figure}
\centering
\fig{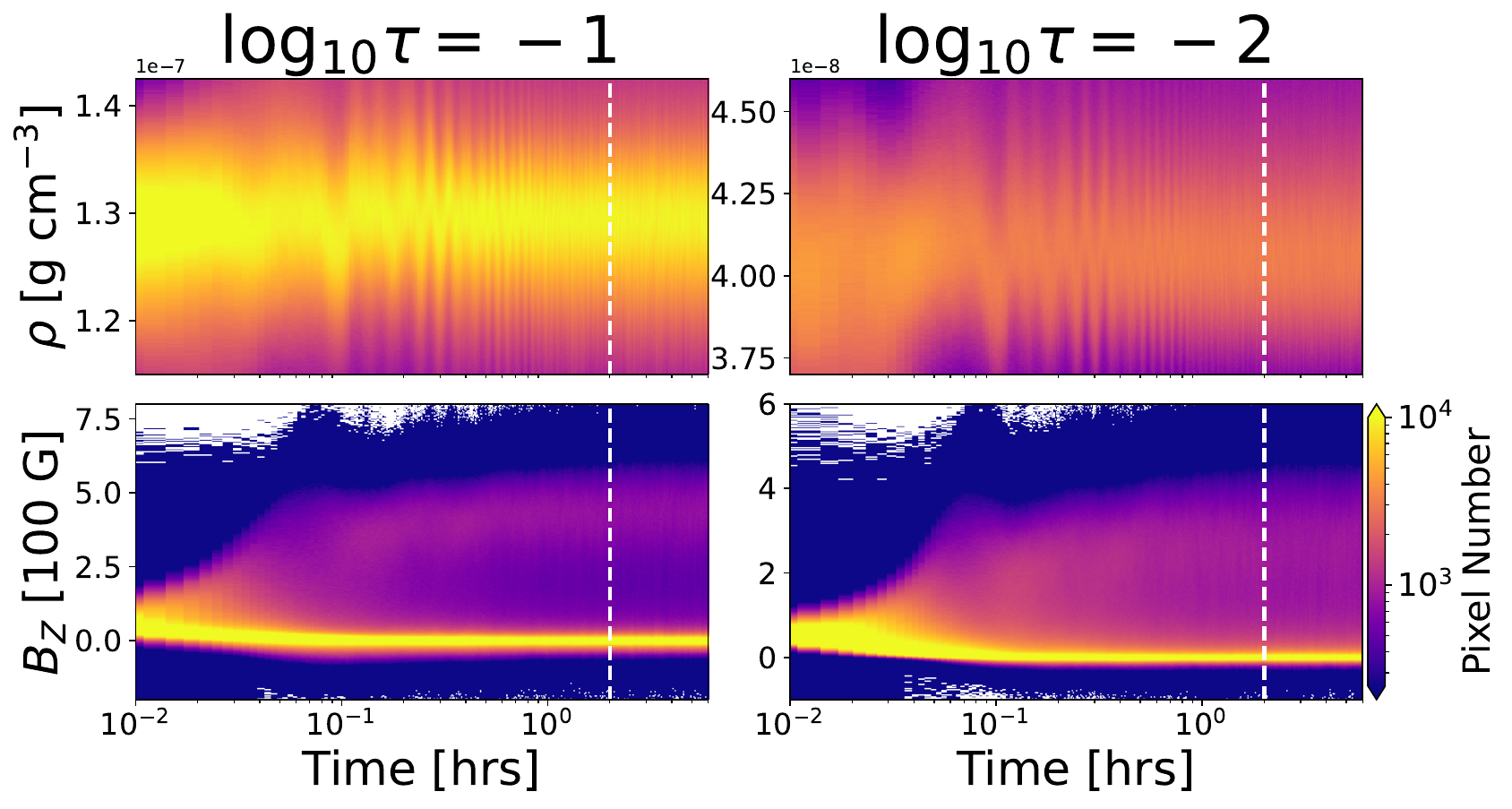}{0.48\textwidth}{}
\caption{The time series histogram of the density, $\rho$, and LOS magnetic field, $B_Z$, on two surfaces, $\log_{10}\tau=-1$ and $-2$, calculated for Case 5. The vertical dashed lines indicate the time of $2$ hrs, after which the simulation has essentially steady state dynamics. \label{fig:4}}
\end{figure}

The addition of vertical fields to the initial conditions of Cases 2--6 represent substantial injections of magnetic energy, and the resulting simulated atmospheres need some time to relax to new steady dynamical states.
To illustrate the atmospheric relaxation process, we focus on Case 5, which features the most intense initially added magnetic field at 200 G. 
\autoref{fig:4} presents time series histograms of physical variables computed at surfaces of $\log_{10}\tau=-1$ and $-2$ in the left and right columns, respectively, i.e., near the maxima of the response functions of the selected \ion{Fe}{1} lines (see \autoref{fig:3}).
The histogram density is displayed as a color scale, with the histogram bin values given on the linear-scaled ordinates, and time given the log-scaled abcissas (to highlight the approach toward a quasi-stationary state). 
Notably, the density $\rho$ distributions exhibit pronounced oscillations attributable to the initial addition of a uniform $B_Z$ and the resulting imbalance of total pressure. The histograms for $B_Z$ exhibits an initial, rapid reduction of pixels with smaller magnitudes (e.g., less than $100$~G), as well as a gradual increase of pixels with larger (e.g., greater than $250$~G).
Both distributions stabilize after approximately two hours (vertical dashed line).

\begin{figure}
\centering
\fig{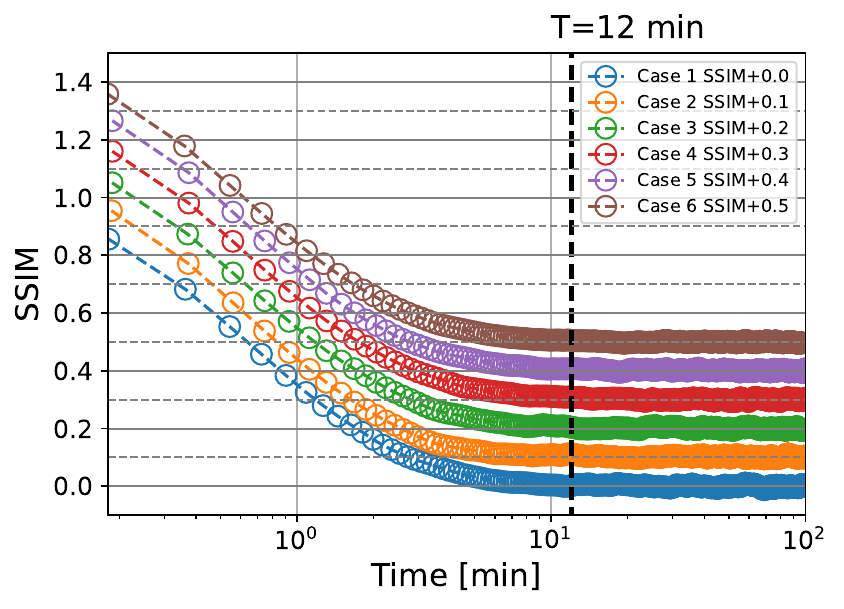}{0.48\textwidth}{}
\caption{The structural similarity index measure (SSIM) of continuum images for different time lag, for all six cases. Each curve is offset vertically by $0.1$ for clarity.  The vertical dashed line marks our selected cadence of 12~minute for producing forward-modeled spectral synthesis of independent observations. \label{fig:5}}
\end{figure}

\begin{figure*}
\centering
\fig{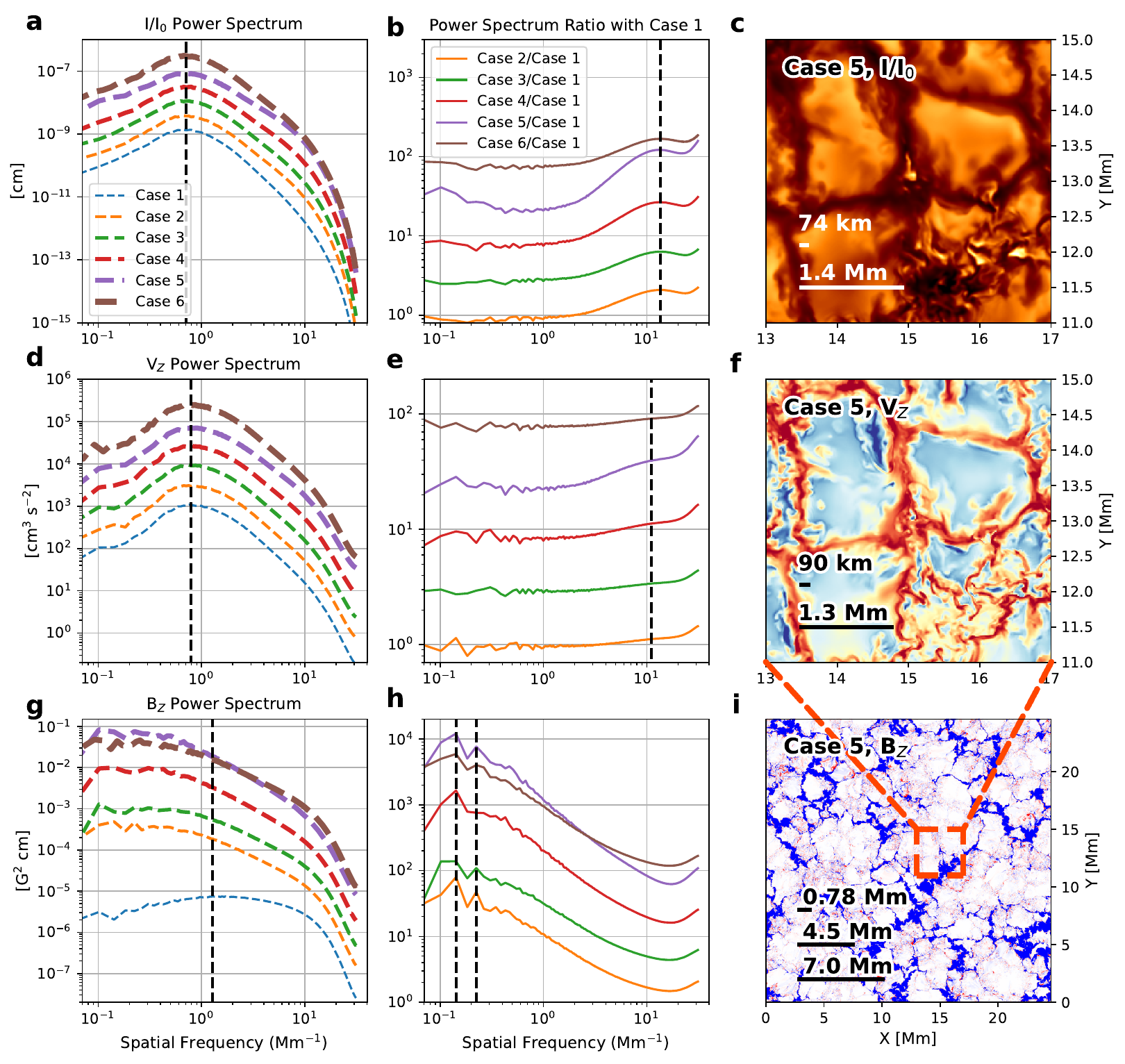}{1\textwidth}{}
\caption{The power spectrum of the normalized continuum image $I/I_0$, $V_Z$, and $B_Z$ at the surface $\log_{10}\tau=0$, displayed from top to bottom. The left panels (a), (d), and (g) present the direct power spectrum for the six cases, each offset vertically by a factor of $3$ for better visualization. The vertical dashed lines in panels (a), (d), and (g) mark the local power spectrum peak for Case 1. The middle panels (b), (e), and (h) depict the power spectrum ratio of Cases 2--6 to Case 1, with vertical dashed lines in (b) and (e) indicating the local peak and saddle point for Case 2 at smaller scales, and in (h), the local peak for Case 2 at larger scales. Panels (c), (f), and (i) show lines marking the typical scales of the peak values in their respective spectra overlaid on a representative sub-region for each variable. The dashed red box in (i) highlights the field of view seen of panels (c) and (f) to enhance the visualization of the typical spatial scale derived from the power spectrum.
\label{fig:6}}
\end{figure*}

With the goal of generating a large training dataset suitable for DL, it is important to ensure that each snapshot of training data contains statistically independent information.
As the primary spatial feature in our simulations is the solar granulation pattern, we seek to minimize the temporal correlation of this pattern in the final training dataset. 
To achieve this, we assess the lifetime of granules in 500 nm continuum intensity by calculating the Structural Similarity Index (SSIM) using the \texttt{skimage}.\texttt{metrics}.\texttt{structural\_similarity} python package \citep{Wang2004ITIP}. 
SSIM returns a scalar value for the difference between two images, where a value of $1 (0)$ indicates identical (different) images.
For each simulation, we calculate SSIM backwards in time, using the final continuum intensity image as a fixed reference image.
The results shown in \autoref{fig:5}. The curve for each case is offset vertically by $0.1$ to aid legibility. Each SSIM curve approaches $0$ after about 10 minutes, indicating significant image variation; such is consistent with a granule's life time of about $10$ minutes. As an initial attempt, we selected a 12-minute cadence to generate the  synthesized the \ion{Fe}{1} lines for our training dataset. More discussion on this choice can be found in Section \ref{sec:3}.

\subsection{On the range of simulated solar environments}

Cases 1--5 represent a range of conditions found on the Sun, differentiated solely by the strength of the uniform magnetic field added to the base SSD simulation for each Case's initial condition; Case 6 extends this with a larger FOV and mixed polarity regions.
The varied conditions produce corresponding differences in the spatial structure of the resulting granulation pattern. 
In \autoref{fig:6}, we quantify those differences by computing, in the $\log_{10}\tau=0$ surface, the spatial power spectra of the normalized continuum intensity $I/I_0$ (row 1, a--c), vertical velocity (row 2, d--f), and vertical magnetic field (row 3, g--i). All the curve plots in the log scale in this figure are offset vertically by multiplying a factor of $3$ for better visualization. 
The effect of adding stronger magnetic fields is made more clear by taking the ratio of the power spectra for Cases 2--6 to the SSD case, as shown in the second column (panels b, d, e).
Together, the power spectra and their ratios reveal that several spatial scales are either present in all cases or notably arise due to the addition of a magnetic field.  
These scales are indicated in the Fourier domain by vertical dashed lines and in the physical domain by solid lines drawn on top of a representative subregion for each variable in panels c, f, and i.

Both the continuum intensity and vertical velocity show pronounced peaks at the scale of the granulation pattern, around 1.4 and 1.3~Mm, as seen in panels (a) and (d), respectively. The vertical magnetic field shows a very slight peak for Case 1 only (panel g); the peaks are indicated by the vertical dashed line in the first column.
From the power spectral ratios presented in the second column, there is a clear trend toward higher power at higher spatial frequencies relative to the SSD case, and a generally steeper slope for cases with a greater added magnetic field strength.
In the continuum intensity, this trend produces the large peak in ratio at a scale of $74$ km, which corresponds to the readily apparent small scale intergranular bright points seen in \autoref{fig:6}(c).
There is no peak in ratio in the power spectra of $V_Z$, but instead a saddle point at a scale of $90$ km (\autoref{fig:6}(e)), which is the width of the intergranular lanes.

Turning our attention to the power spectrum of $B_Z$, the only apparent peak occurs for Case 1 and has size $0.78$ Mm, i.e., roughly half of the granule size as measured in either continuum intensity or velocity. 
Surprisingly, as more flux is added to the simulation the magnetic field becomes more ordered (more distinct local peaks) on larger scales.
This somewhat mimics a mesogranular morphology and has two local peaks at $4.5$ and $7.0$ Mm (\autoref{fig:6}(h,i)).
However, this might be influenced by interaction with bottom boundary, since we do not have a deep enough convection layer for the formation of the supergranulation convection pattern \citep{Lord2014ApJ,lord2014b}.

\begin{figure*}
\centering
\fig{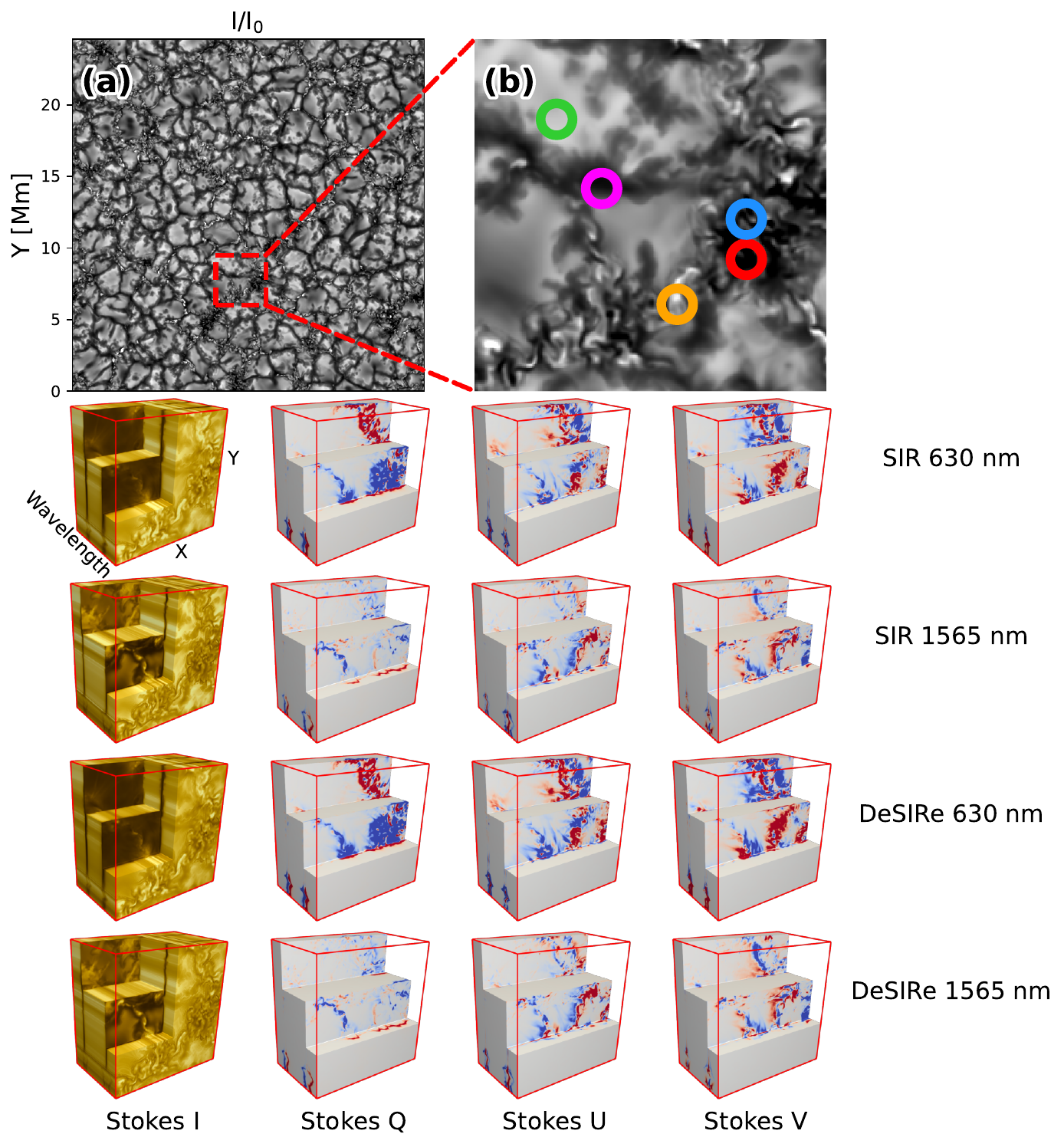}{\textwidth}{}
\caption{Example synthesized spectropolarimetric data computed for a single snapshot of the Case 5 MHD simulation at $t=5.6$ hrs, 3D data cubes show a subset of the output for both spectral line pairs and both forward modeling codes. The red box in panel (a) marks the field of view of the 3D data cubes. Panel (b) shows the zoom-in view of the subset with colorized circles indicating locations representative of typical Stokes profiles, as presented in \autoref{fig:10}. \label{fig:7}}
\end{figure*}



\section{Stokes Profile Synthesis}\label{sec:3}

\begin{deluxetable}{ccc}
\tablenum{2}
\tablecaption{Stokes Profile Data \label{tab:2}}
\tablewidth{0pt}
\tablehead{
\colhead{\ion{Fe}{1}} & \colhead{6301-6302 \AA \ } & \colhead{15648-15652 \AA \ }
}
\startdata
Line Range (\AA) & 6300.8--6303.3 & 15646.8--15654.8 \\
$\Delta\lambda$ (m\AA) & 8.945 & 31.376 \\
SIR Data Size (TB) & 5.5 & 5.1 \\
DeSIRe Data Size (TB) & 5.5 & 5.1
\enddata
\end{deluxetable}



We calculated synthetic Stokes data for each Case at the statistically independent 12 minute cadence after the initial 2 hours relaxation period, as determined in Section~\ref{sec:2}. 
For the Stokes profiles dataset, we selected two \ion{Fe}{1} line pairs in the $630$~nm and $1565$ nm range, originating from the deep and upper layers of the photosphere, respectively. These lines, whose response functions are depicted in \autoref{fig:3}(a)--(d), are pivotal in diagnosing the photospheric magnetic field \citep[see][and reference therein]{BellotRubio2019LRSP}. Employing a multi-line diagnostic approach with these lines will facilitate a comprehensive understanding of the three-dimensional photospheric structure.

\begin{deluxetable*}{ccccccc}
\tablenum{3}
\tablecaption{Atomic Parameters \label{tab:3}}
\tablewidth{0pt}
\tablehead{
\colhead{Wave length (\AA)} & \colhead{$\Gamma_6$\tablenotemark{a}} & \colhead{Exc. Pot \tablenotemark{b} (eV)} & \colhead{log($gf$)\tablenotemark{c}} & \colhead{Transition Level} & \colhead{$\alpha$\tablenotemark{d} } & \colhead{$\sigma$\tablenotemark{d} (cm$^2$) }}
\startdata
6301.5012     &  1.0 & 3.654  &-0.718  &  5P 2.0- 5D 2.0                             & 0.242  & 2.33543e-14 \\
6302.4936     &  1.0 & 3.686  &-1.131  &  5P 1.0- 5D 0.0                             & 0.239  & 2.38024e-14  \\
15648.5088   &  1.0 & 5.426  & -0.652 &  7D 1.0- 7D 1.0                             & 0.229  & 2.72747e-14 \\
15652.8809   &  1.0 & 6.246  & -0.050 & 7D 5.0- 7k 4.0\tablenotemark{e}   & 0.330  & 4.045e-14   
\enddata
\tablecomments{
The atomic parameters are written in SIR format \citep{RuizCobo1992}.
\tablenotetext{a}{Enhancement factor to the van der Waals coefficient.}
\tablenotetext{b}{Excitation potential of the lower level.}
\tablenotetext{c}{The logarithm of the multiplicity of the level times the oscillator strength.}
\tablenotetext{d}{The collisional broadening parameters from the quantum mechanical theory of \cite{Barklem1998PASA}.}
\tablenotetext{e}{This transition level is used for SIR synthesis pipeline, while for the DeSIRe, the level is 7D 5.0- (6D4.5)f2k 4.0.}}
\end{deluxetable*}

\begin{figure*}
\centering
\fig{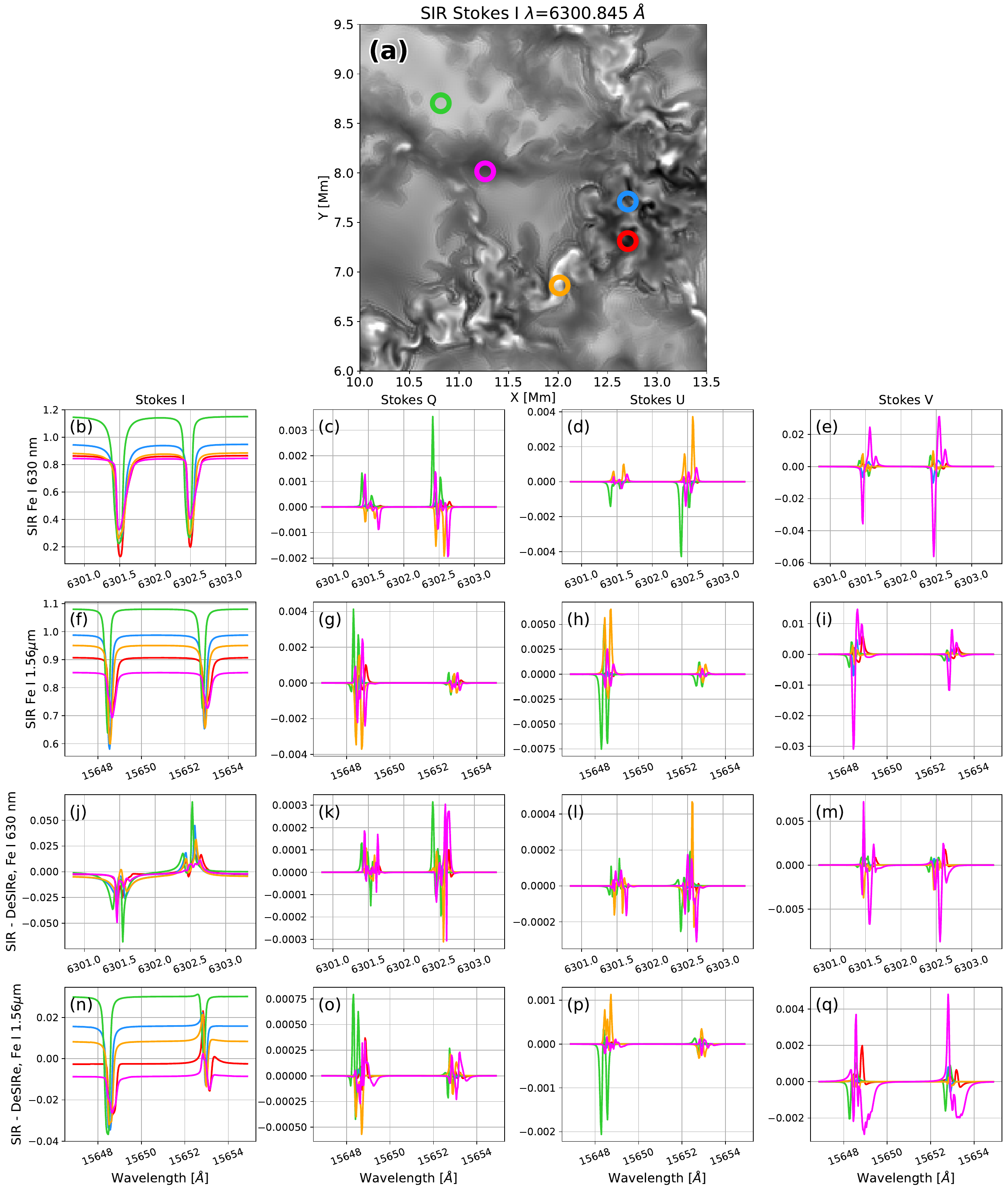}{0.9\textwidth}{}
\caption{Illustration of the synthesized Stokes profiles from SIR. (a) Intensity of SIR Stokes $I$ at continuum wavelength, with locations of representative Stokes profiles marked by colored circles. Panels (b)-(e) and (f)-(i) show the Stokes profiles for the $630$ nm and $1565$ nm lines, respectively. Panels (j)-(m) and (n)-(q) detail the differences between the SIR and DeSIRe synthesized profiles, i.e., $I_\mathrm{SIR} - I_\mathrm{DeSIRe}$. \label{fig:10}}
\end{figure*}


\begin{figure*}
\centering
\fig{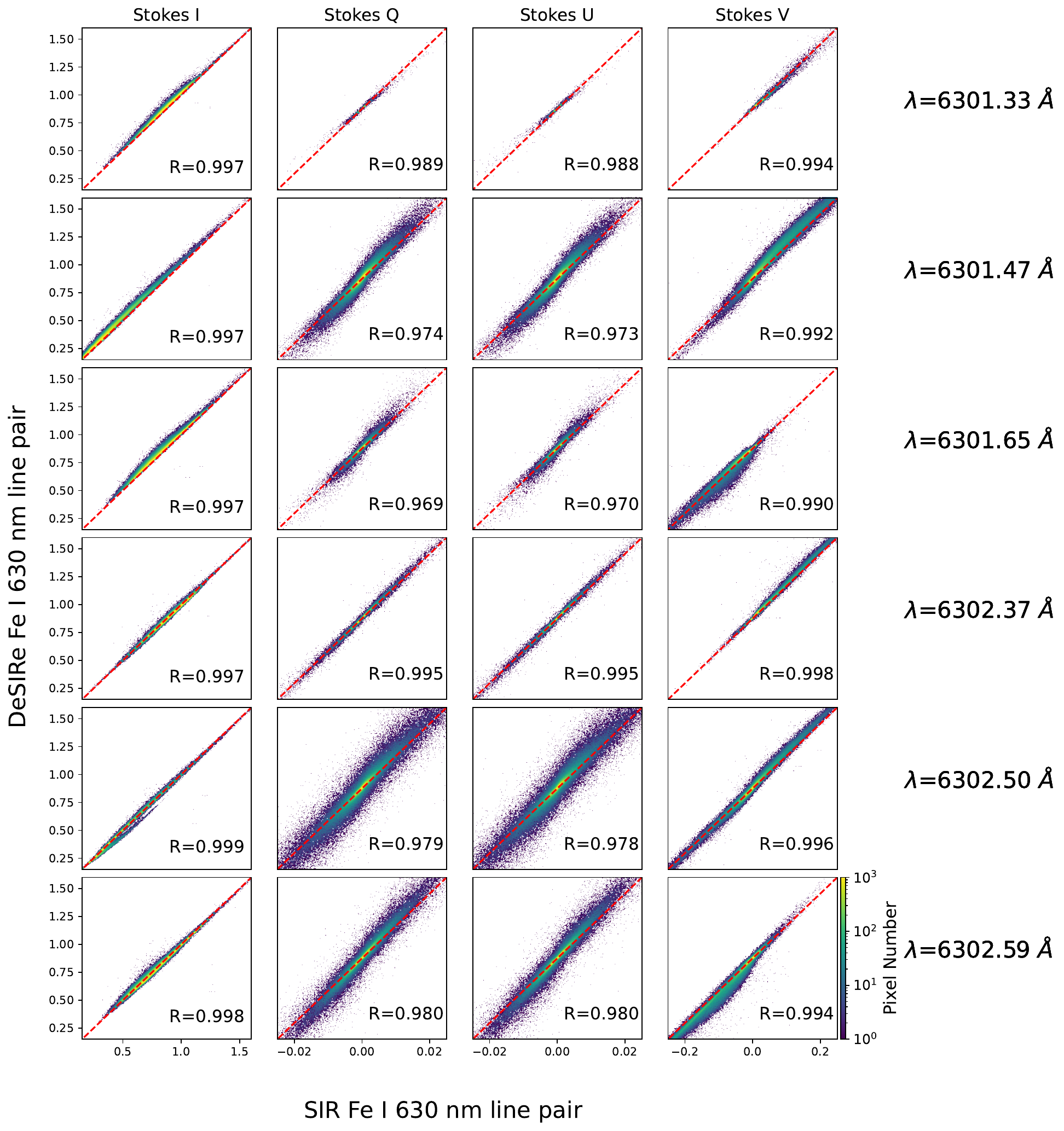}{\textwidth}{}
\caption{Comparison of the Stokes data from SIR and DeSIRe for the 630 nm line from Case 5 with $t=5.6$ hrs, focusing on the wings and cores. The colors represent the pixel count in the 2D histogram. Dashed red lines denote the line of identity, indicating perfect agreement between the two sets of data. The corresponding Pearson correlation coefficients are marked on each panel. \label{fig:8}}
\end{figure*}

The output wavelength sampling for the synthesis was chosen to match the capabilities of the DL-NIRSP instrument, as detailed in \citet{Jaeggli2022SoPh}. For the $630$~nm spectral window, the DL-NIRSP has a nominal bandpass (derived from the combination of a narrowband filter and spectral mask) of $6.4$ \AA, \ a spectral sampling rate of approximately $17.9$ m\AA \ per pixel, and a point spread function (PSF) with a full width at half maximum (FWHM) of about $40$ m\AA. Our synthesis was performed at half the DL-NIRSP sample step, about $8.945$ m\AA, covering a range from $6300.8521$ \AA \ to $6303.3119$ \AA \ across $275$ steps.
These settings fully cover both lines and extend well into the continuum on either side.

Similarly, for the $1565$ nm \ion{Fe}{1} line pair, the DL-NIRSP as a nominal bandpass off $16.1$ \AA, a spectral sampling rate of $62.9$ m\AA \ per pixel, and a PSF with FWHM of $125.8$ m\AA. We again synthesized at double the DL-NIRSP spectral sampling, $31.376$ m\AA, covering a wavelength range of $15646.875$ \AA \ to $15654.876$ \AA \ with $255$ steps. The settings for both line pairs are detailed in Table \ref{tab:2}.

We performed the spectral synthesis using two different codes.
First, for forward synthesis under local thermal equilibrium (LTE) conditions, we used the 3D version of SIR \citep{RuizCobo1992,asensioramos2019}. 
To align with our dataset and coordinate system, the code underwent several adaptations, with the revised version accessible on GitHub\footnote{\url{https://github.com/ifauh/par-sir}} and details of the modifications described in \autoref{sec:append0}. Acknowledging the significance of non-LTE effects as identified in the research by \cite{Smitha2020AA,Smitha2021AA}, we also integrated the DeSIRe code \citep{RuizCobo2022}, along with the parallel Python wrapper outlined by \cite{Gafeira2021AA}, into our software pipeline. This approach facilitated the generation of non-LTE Stokes profiles for our chosen lines. 
The elemental abundance data for SIR source is from \cite{Grevesse1998SSRv}, which is consistent with that used in MURaM simulation. For DeSIRe we use an updated abundance data from \cite{Asplund2009ARAA}. The atomic parameters for the \ion{Fe}{1} lines are detailed in Table \ref{tab:3}. The resulting synthesized spectral datasets are illustrated in \autoref{fig:7}.


Cases 2--5 are characterized by a dominant positive magnetic flux. To account for the equally possible input of negative polarity, we manually flipped the signs of the magnetic field, and calculate a second set of synthetic profiles for each dataset. Note that the plasma evolution under the ideal MHD equations remains unchanged when the sign of all three magnetic field components are inverted.
This approach allows for random selection between the original and flipped magnetic fields during training of the machine learning model, effectively minimizing the possible bias from the magnetic field polarity in the resulting neural network model.
In total, we generated a collection of $20$ ~TB of Stokes profiles for both SIR and DeSIRe, with and without magnetic field flipping, covering $210$ dynamically independent snapshots. 

Stokes profiles for five representative locations are shown in \autoref{fig:10}, corresponding to Case 5 at $t=5.6$ hrs. The first (b--e) and second (f--i) rows present the Stokes profiles for \ion{Fe}{1} $630$ nm and $1565$ nm line pairs from the SIR code. The third (j--m) and fourth (m--q) rows present the difference between the synthesized profiles from SIR and DeSIRe, demonstrating minimal discrepancies (note the scale difference between the upper and lower rows).
\autoref{fig:8} presents the joint distribution between the SIR and DeSIRe synthesized Stokes profiles for the $630$ nm line pair for the same dataset, with the SIR results shown on the abscissa and DeSIRe on the ordinate for each panel.  The red dashed curve marks the one-to-one line.  From top to bottom, the rows compare wavelengths in the blue wing, line core, and red wing of the $630.15$ nm line, followed by similar plots for the $630.25$ nm line. The Pearson correlation of the Stokes I and V is always greater than $0.99$, while the linear polarizations (Q and U) show a slightly smaller correlation, $R\approx 0.96$. \autoref{fig:a0} shows the analogous comparison for the $1565$ nm line pair, in \autoref{sec:append0}. The high correlation suggests that the results from the two synthesis codes are in qualitative agreement. The differences can be owing in part to the LTE versus non-LTE treatment of line formation or the different opacity packages.

\begin{figure*}
\centering
\fig{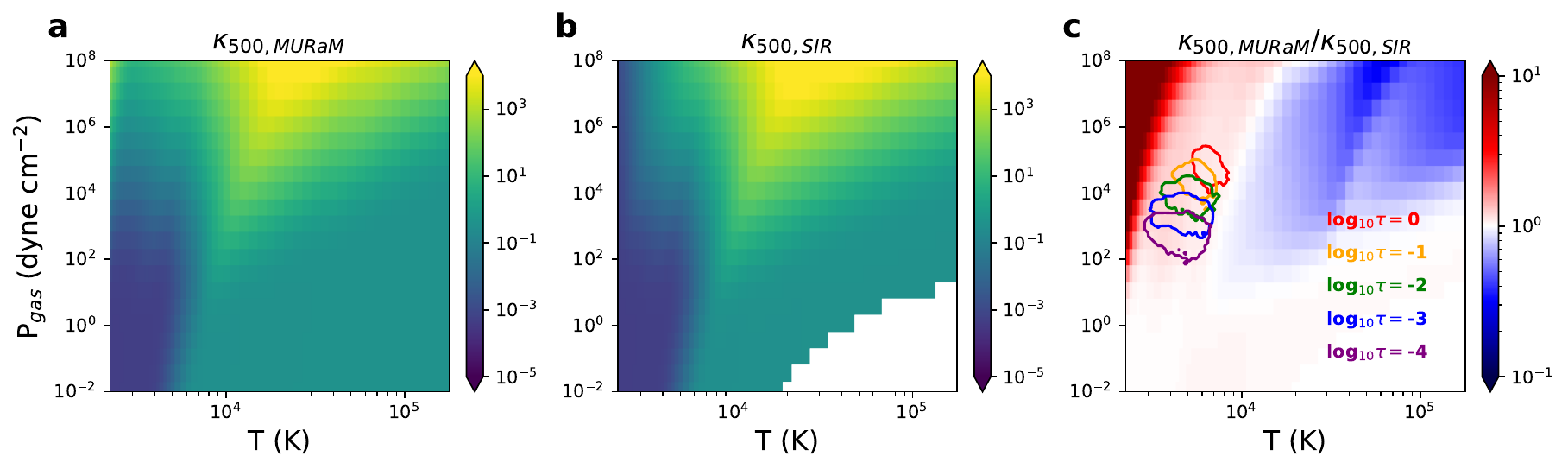}{\textwidth}{}
\caption{Panels (a) and (b) display the continuum opacity tables derived from MURaM and SIR models, respectively. Panel (c) illustrates the ratio of these opacity tables, highlighting the pressure and temperature ranges with colored lines for Case 1 at $t=5.4$ hrs across various $\log_{10}\tau$ layers.\label{fig:12}}
\end{figure*}

\begin{figure*}
\centering
\fig{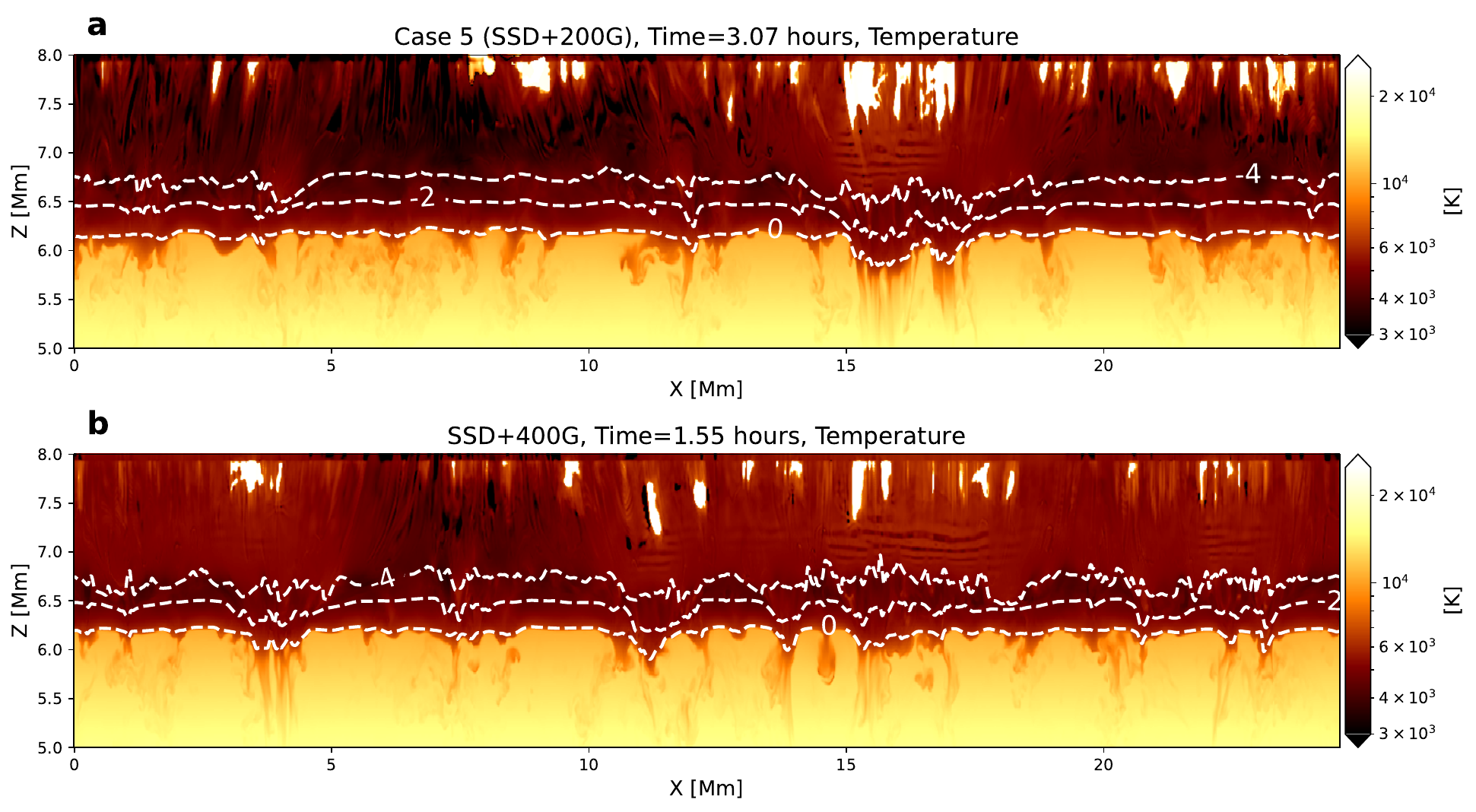}{\textwidth}{}
\caption{The oscillation in temperature in the vertical slice in the middle of the simulation box. Panel (a) and (b) show that from the Case 5 and an additional test case with an added uniform field be as large as 400 G. The three dotted curves are for $\log_10\tau=0$, $-2$, and $-4$.\label{fig:11}}
\end{figure*}

Our analysis revealed discrepancies in the continuum opacity as a function of $T$ and gas pressure $P_\text{gas}$ between the SIR and MURaM results. 
\autoref{fig:12} contrasts their $\kappa_{500}$ values. MURaM's values are derived from its own lookup tables \citep{Castelli2004A&A...419..725C}, while SIR's are calculated with its \texttt{PemufromPgT\_i} module using $T$ and $P_\text{gas}$. Both use the same element abundance \citep{Grevesse1998SSRv}. 
Notably, in certain optical depth layers where the line is formed (outlined by the colored contours in \autoref{fig:12}(c)), the opacity from MURaM is slightly larger than that from SIR. 

A direction comparison between MURaM and DeSIRe opacity prove to be difficult. The latter uses a different opacity package from the RH code \citep{Uitenbroek2001ApJ...557..389U}. It is calculated under non-LTE conditions, which does not solely depend on local thermal variables. This is expected to cause some difference in the synthetic profiles shown in Figures \ref{fig:10}, \ref{fig:8}, and \ref{fig:a0}. As it is out of the scope of the SPIn4D project, we defer a comprehensive investigation of their differences to future work.
\section{Artifacts in MHD Simulation and Solutions}\label{sec:simulationproblems}


Several features of the MHD simulations may require attention when using the simulation database. 
In all six cases, significant oscillations occur in the upper regions of the simulated atmosphere, close to the top boundary. These are especially apparent for Case 5, with the strongest magnetic field of $200$~G (\autoref{fig:11}(a)). 
The root cause for this is that we artificially limited the Alfv\'en velocity $v_A$ to $60$~km~s$^{-1}$ to speed up the computation \citep{Rempel2009ApJ,rempel2017} .
This is justified for our primary target, the photosphere, where $v_A$ is low. However, numerical issues may arise in higher atmospheric layers as $v_A$ increases rapidly with decrease density such that the imposed limit becomes inconsistent with the calculation.
To address this issue, we adjusted the Courant-Friedrichs-Lewy (CFL) condition and Alfv\'en velocity limit as necessary to maintain numerical stability.
The approach proves to significantly reduce the oscillation, though some residual signals remain. Other solutions include adding proper treatment for a transition region in higher layers. Unfortunately they would lead to more than a tenfold increase in the required computing time, so were not adopted there.

\revision{It is worth noting that such oscillations do not appear in \cite{rempel2014},  and they} are barely detectable \revision{for the cases with weaker mean magnetic field.} Specifically, Case 5 is intermittently impacted, whereas Case 4 remains mostly unaffected. As a test, we also simulated a case with a stronger, $400$~G added magnetic field, which produced even stronger oscillations (\autoref{fig:11}(b)). \revision{The problem arises from both stronger magnetic fields and the more extended vertical domain. The combination leads to large $v_A$ values exceeding the imposed limit.} Efforts to stabilize this simulation by reducing the CFL parameter and adjusting velocity parameters were unsuccessful. A potential solution involves expanding the simulation box upwards and incorporating a realistic corona. This would significantly increase computational demands, exceeding the SPIn4D project's scope. Consequently, we limited our simulations to a maximum of $200$~G magnetic field strength.






In both 200 G and 400 G cases, we observed numerous ``hot pockets" in the regions close to the top boundaries.  
In these regions, the code attempts to form a hot layer analogous to the transition region, but the set boundary conditions do not allow it to do so.
These hot pockets tend to extend to the lower region as the mean magnetic field increases. Meanwhile, temperatures outside these pockets above $z>7$~Mm rise slightly with a stronger mean magnetic field, though the pockets themselves show little correlation with changes in the magnetic field.

We note that these aforementioned artifacts are not expected to impact our results, as illustrated by \autoref{fig:11}.
The upper panel (a) is for Case 5.
The regions of numerical instability produce very high temperatures, exceeding $2\times10^4$ K, saturating the color table in white. 
These regions produce both the strong oscillations and shocks in the surrounding plasma, but are basically confined to strong magnetic field regions.
\revision{Since the MURaM code uses a numerical diffusivity scheme that can significantly enhance diffusivities in regions where monotonicity changes \citep[see Sect. 2 in][]{rempel2014} where needed and keep them low everywhere else. This sufficiently decays and prevents the spread of these oscillations to the lower and higher atmosphere, mainly trapped around $z=7$ Mm where they are generated.}
The magnitude of the oscillations decreases as the density rises in the lower regions of the atmosphere, such that the oscillations are largely confined to layers where $\log_{10}\tau<-4$.
Given the distribution of the response function, as shown in \autoref{fig:3}(a)--(d), the regions critical for the formation of \ion{Fe}{1} lines remain unaffected.



\section{Summary}\label{sec:4}

In this work, we provide an overview of the SPIn4D project, which aims at advancing the inversion of spectropolarimetric data through machine learning. We describe the procedures for generating a comprehensive training and test dataset derived from MURaM simulations and the forward synthesis of Stokes profiles. Specifically, we conducted six distinct MURaM SSD simulations, generating a total of 109 TB of photospheric atmosphere data. Additionally, we synthesized \ion{Fe}{1} lines at $630$ nm and $1565$ nm for every $12$-minute from the simulations, yielding 21 TB of data in HDF5 format. 
The simulations required an extensive computational effort, amounting to 10 million CPU hours. We have released the SIR-based synthesized \ion{Fe}{1} lines and the corresponding 3D photospheric slabs totaling 13.7 TB, making them accessible to the wider research community for further analysis and study.
We have used both SIR and DeSIRe codes to synthesize the Stokes profiles. Their results largely agree; additional study is required to explain their minor differences. We will focus on the SIR results and keep the DeSIRe version available for the community.
Our DL model training will use these MURaM simulations. The results and comparisons with the SIR inversion as a baseline will be presented in upcoming work.

This dataset, encompassing photospheric physical variables from both quiet Sun and Plage regions, is poised to bolster the burgeoning field of machine learning within solar physics. Its relevance extends particularly to the early research topics of the \textit{DKIST} science objectives \citep{rast2021}. The versatility of the data not only supports the inversion tasks of the SPIn4D project but also potentially facilitates the development of a range of other ML models. The public availability of this dataset ensures that the broader scientific community can leverage it not just for inversion studies but also for other tasks, like advanced super-resolution and disambiguation of the horizontal magnetic field. The large volume can reduce the issue of overfitting when training DL models on a small dataset. Its applicability is not limited to \textit{DKIST} alone; other solar telescopes can also benefit from the insights derived from this comprehensive dataset and the ML model built on it. In this way, the dataset acts as a critical resource, driving forward the integration of machine learning techniques into solar physics and potentially transforming observational strategies and data analysis methodologies across multiple platforms.

\begin{acknowledgments}
We thank Carlos Quintero Noda for helpful discussions. This work is supported by NSF/AAG award \#2008344, NSF/CAREER award \#1848250 to the University of Hawai`i (UH), and the state of Hawai`i. J.L. acknowledges the support of the \textit{DKIST} Ambassador program. This material is based upon work supported by the NSF National Center for Atmospheric Research, which is a major facility sponsored by the U.S. National Science Foundation under Cooperative Agreement No. 1852977. We would like to acknowledge high-performance computing support from \textit{Cheyenne} (doi: \href{https://dx.doi.org/10.5065/D6RX99HX}{10.5065/D6RX99HX}) provided by NSF NCAR's Computational and Information Systems Laboratory (CISL), sponsored by the NSF. The technical support and advanced computing resources from UH Information Technology Services -- Cyberinfrastructure, funded in part by the NSF CC* awards \#2201428 and \#2232862 are gratefully acknowledged.
L.A.T, S.A.J., and T.A.S are supported by the National Solar Observatory.
\end{acknowledgments}

\vspace{5mm}
\facilities{Casper and Cheyenne HPC at NCAR, Mana and Koa HPC at the University of Hawai`i at M\=anoa.}

\software{MURaM \citep{Vogler2005AA,Rempel2009ApJ}, SIR3D \citep{RuizCobo1992,ruizcobo2012}, DeSIRe \citep{RuizCobo2022}, Parallel Python wrapper \citep{Gafeira2021AA}, Matplotlib \citep{Hunter2007}, and SciPy \citep{2020SciPy-NMeth}.}
          

\appendix

\section{Changes in SIR3D and Data Organizing}\label{sec:append0}

To enhance the handling of half-integer quantum levels for the \ion{Fe}{1} $1565$~nm lines, new atomic state symbols have been updated in the \texttt{src/interface.f90} module of the 3D version of SIR. The orbital angular momentum range has been expanded from six integer levels to include 13 integer and 11 half-integer levels. The element abundance data was updated in both \texttt{src/interface.f90} and \texttt{src/leyendo.f}. We have enhanced the selection of optical depth ranges for forward synthesis by introducing new parameters in \texttt{synth/model.py}: \texttt{clip\_tau}, \texttt{clip\_tau\_min}, and \texttt{clip\_tau\_max}. To address precision issues in the upper layers of MURaM simulations that cause duplication at very small optical depths, line synthesis is now constrained to layers within $-5 < \log_{10}\tau < 2$.
We modified the SIR3D code to support a new `MURAM' atmosphere that uses solar Cartesian coordinates to match the DL-NIRSP axis ordering and the conventions used in the \textit{Python} package for the solar community (SunPy).
MURaM's SSD simulation coordinates ($z^m$, $y^m$, $x^m$) correspond to the SIR3D's coordinates ($z^s$, $x^s$, $y^s$) and to solar Cartesian coordinates ($y^c$, $x^c$, $z^c$), thus a \texttt{transpose(1,0,2)} transformation is applied to all \texttt{memmap} objects in \texttt{synth/multiprocessing.py} to align with the solar Cartesian coordinate system.
See our project website for a detailed explanation of these changes.

On the SPIn4D project website, our datasets are stored within corresponding case directories, whose names are characterized by the mean magnetic field strength. The directories are labeled as follows: \texttt{SPIN4D\_SSD}, \texttt{SPIN4D\_SSD\_50G}, \texttt{SPIN4D\_SSD\_50G\_V}, \texttt{SPIN4D\_SSD\_100G}, \texttt{SPIN4D\_SSD\_200G}, and \texttt{SPIN4D\_SSD\_Large}, corresponding to Cases 1 through 6. For instance, the file \texttt{subdomain\_0.051405} in the directory \texttt{SPIN4D\_SSD\_50G} corresponds to the mass density data from the simulation output at time index \texttt{"051405"}, where \texttt{"0"} is the variable index. Variable indices from \texttt{0} to \texttt{11} represent mass density ($\rho$), velocities ($v_z$, $v_x$, $v_y$), internal energy ($e$), magnetic fields ($B_z$, $B_x$, $B_y$), temperature ($T$), pressure ($P$), electron number density ($N_e$), and optical depth ($\tau$), the shift of the vector components reflecting the coordinate transformations between SIR and MURaM aforementioned. Furthermore, \texttt{stokes-051405-6302.h5} corresponds to the \ion{Fe}{1} 630 nm line at the corresponding output time index. Detailed guidance on accessing and interpreting the data is provided in our online tutorial\footnote{\url{https://github.com/ifauh/spin4d-data}}. \revision{It is also available on Zenodo \dataset[doi:10.5281/zenodo.13879854]{https://doi.org/10.5281/zenodo.13879854}.}

\section{Comparison of Synthesized Lines} \label{sec:append1}

\begin{figure}
\centering
\fig{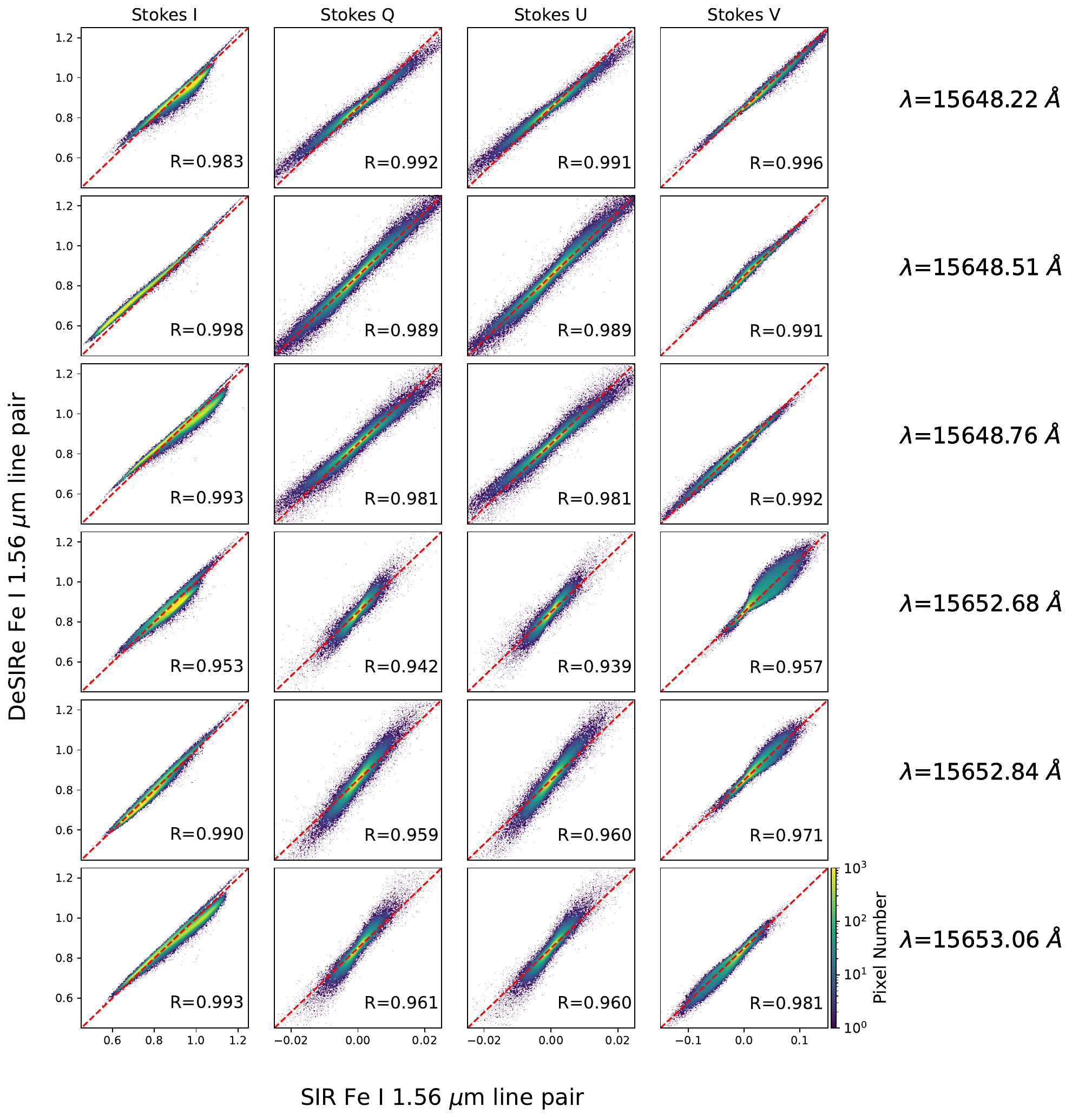}{\textwidth}{}
\caption{The same as \autoref{fig:8} but for the $1565$ nm lines.\label{fig:a0}}
\end{figure}

Statistical comparisons of the synthesized Stokes profiles for the $1565$~nm line from SIR and DeSIRe are detailed in \autoref{fig:a0}. These discrepancies of the Stokes $I$ and $V$ are more obvious than those observed in the $630$~nm lines shown in \autoref{fig:8}. Specifically, for the $1565.2$~nm line, DeSIRe exhibits greater slopes for Stokes $Q$ and $U$ compared to the identity relation. Despite these differences, the Pearson correlation coefficient remains above $0.939$ for all line positions and all Stokes components, though it shows a weaker correlation than the $630$~nm lines, whose smallest correlation is $0.969$.



\end{CJK*}

\bibliographystyle{aasjournal}

\begin{thebibliography}{108}
\expandafter\ifx\csname natexlab\endcsname\relax\def\natexlab#1{#1}\fi

\bibitem[{{Antiochos}(1998)}]{Antiochos1998ApJ}
{Antiochos}, S.~K. 1998,
  \href{http://dx.doi.org/10.1086/311507}{\color{magenta}\apjl},
  \href{https://ui.adsabs.harvard.edu/abs/1998ApJ...502L.181A}{502, L181}

\bibitem[{{Antiochos} {et~al.}(1999){Antiochos}, {DeVore}, \&
  {Klimchuk}}]{Antiochos1999ApJ}
{Antiochos}, S.~K., {DeVore}, C.~R., \& {Klimchuk}, J.~A. 1999,
  \href{http://dx.doi.org/10.1086/306563}{\color{magenta}\apj},
  \href{https://ui.adsabs.harvard.edu/abs/1999ApJ...510..485A}{510, 485}

\bibitem[{{Asensio Ramos} {et~al.}(2023){Asensio Ramos}, {Cheung}, {Chifu}, \&
  {Gafeira}}]{AsensioRamos2023LRSP}
{Asensio Ramos}, A., {Cheung}, M. C.~M., {Chifu}, I., \& {Gafeira}, R. 2023,
  \href{http://dx.doi.org/10.1007/s41116-023-00038-x}{\color{magenta}Living
  Reviews in Solar Physics},
  \href{https://ui.adsabs.harvard.edu/abs/2023LRSP...20....4A}{20, 4}

\bibitem[{{Asensio Ramos} {et~al.}(2018){Asensio Ramos}, {de la Cruz
  Rodr{\'\i}guez}, \& {Pastor Yabar}}]{AsensioRamos2018A&A...620A..73A}
{Asensio Ramos}, A., {de la Cruz Rodr{\'\i}guez}, J., \& {Pastor Yabar}, A.
  2018,
  \href{http://dx.doi.org/10.1051/0004-6361/201833648}{\color{magenta}\aap},
  \href{https://ui.adsabs.harvard.edu/abs/2018A&A...620A..73A}{620, A73}

\bibitem[{{Asensio Ramos} \& {D{\'\i}az Baso}(2019)}]{asensioramos2019}
{Asensio Ramos}, A. \& {D{\'\i}az Baso}, C.~J. 2019,
  \href{http://dx.doi.org/10.1051/0004-6361/201935628}{\color{magenta}Astron.
  Astrophys.},
  \href{https://ui.adsabs.harvard.edu/abs/2019A&A...626A.102A}{626, A102}

\bibitem[{{Asensio Ramos} {et~al.}(2017){Asensio Ramos}, {Requerey}, \&
  {Vitas}}]{asensioramos2017}
{Asensio Ramos}, A., {Requerey}, I.~S., \& {Vitas}, N. 2017,
  \href{http://dx.doi.org/10.1051/0004-6361/201730783}{\color{magenta}Astron.
  Astrophys.},
  \href{https://ui.adsabs.harvard.edu/abs/2017A&A...604A..11A}{604, A11}

\bibitem[{{Asensio Ramos} {et~al.}(2008){Asensio Ramos}, {Trujillo Bueno}, \&
  {Landi Degl'Innocenti}}]{AsensioRamos2008ApJ}
{Asensio Ramos}, A., {Trujillo Bueno}, J., \& {Landi Degl'Innocenti}, E. 2008,
  \href{http://dx.doi.org/10.1086/589433}{\color{magenta}\apj},
  \href{https://ui.adsabs.harvard.edu/abs/2008ApJ...683..542A}{683, 542}

\bibitem[{{Asplund} {et~al.}(2009){Asplund}, {Grevesse}, {Sauval}, \&
  {Scott}}]{Asplund2009ARAA}
{Asplund}, M., {Grevesse}, N., {Sauval}, A.~J., \& {Scott}, P. 2009,
  \href{http://dx.doi.org/10.1146/annurev.astro.46.060407.145222}{\color{magenta}\araa},
  \href{https://ui.adsabs.harvard.edu/abs/2009ARA&A..47..481A}{47, 481}

\bibitem[{{Barklem} {et~al.}(1998){Barklem}, {Anstee}, \&
  {O'Mara}}]{Barklem1998PASA}
{Barklem}, P.~S., {Anstee}, S.~D., \& {O'Mara}, B.~J. 1998,
  \href{http://dx.doi.org/10.1071/AS98336}{\color{magenta}\pasa},
  \href{https://ui.adsabs.harvard.edu/abs/1998PASA...15..336B}{15, 336}

\bibitem[{{Bellot Rubio} \& {Orozco Su{\'a}rez}(2019)}]{BellotRubio2019LRSP}
{Bellot Rubio}, L. \& {Orozco Su{\'a}rez}, D. 2019,
  \href{http://dx.doi.org/10.1007/s41116-018-0017-1}{\color{magenta}Living
  Reviews in Solar Physics},
  \href{https://ui.adsabs.harvard.edu/abs/2019LRSP...16....1B}{16, 1}

\bibitem[{{Bobra} \& {Couvidat}(2015)}]{bobra2015}
{Bobra}, M.~G. \& {Couvidat}, S. 2015,
  \href{http://dx.doi.org/10.1088/0004-637X/798/2/135}{\color{magenta}Astrophys.
  J.}, \href{http://adsabs.harvard.edu/abs/2015ApJ...798..135B}{798, 135}

\bibitem[{{Bobra} \& {Ilonidis}(2016)}]{bobra2016}
{Bobra}, M.~G. \& {Ilonidis}, S. 2016,
  \href{http://dx.doi.org/10.3847/0004-637X/821/2/127}{\color{magenta}Astrophys.
  J.}, \href{https://ui.adsabs.harvard.edu/abs/2016ApJ...821..127B}{821, 127}

\bibitem[{{Borrero} \& {Pastor Yabar}(2023)}]{Borrero2023AA}
{Borrero}, J.~M. \& {Pastor Yabar}, A. 2023,
  \href{http://dx.doi.org/10.1051/0004-6361/202244716}{\color{magenta}\aap},
  \href{https://ui.adsabs.harvard.edu/abs/2023A&A...669A.122B}{669, A122}

\bibitem[{{Borrero} {et~al.}(2019){Borrero}, {Pastor Yabar}, {Rempel}, \& {Ruiz
  Cobo}}]{Borrero2019AA}
{Borrero}, J.~M., {Pastor Yabar}, A., {Rempel}, M., \& {Ruiz Cobo}, B. 2019,
  \href{http://dx.doi.org/10.1051/0004-6361/201936367}{\color{magenta}\aap},
  \href{https://ui.adsabs.harvard.edu/abs/2019A&A...632A.111B}{632, A111}

\bibitem[{{Borrero} {et~al.}(2021){Borrero}, {Pastor Yabar}, \& {Ruiz
  Cobo}}]{Borrero2021AA}
{Borrero}, J.~M., {Pastor Yabar}, A., \& {Ruiz Cobo}, B. 2021,
  \href{http://dx.doi.org/10.1051/0004-6361/202039927}{\color{magenta}\aap},
  \href{https://ui.adsabs.harvard.edu/abs/2021A&A...647A.190B}{647, A190}

\bibitem[{{Brehmer} {et~al.}(2020){Brehmer}, {Louppe}, {Pavez}, \&
  {Cranmer}}]{Brehmer2020PNAS}
{Brehmer}, J., {Louppe}, G., {Pavez}, J., \& {Cranmer}, K. 2020,
  \href{http://dx.doi.org/10.1073/pnas.1915980117}{\color{magenta}Proceedings
  of the National Academy of Science},
  \href{https://ui.adsabs.harvard.edu/abs/2020PNAS..117.5242B}{117, 5242}

\bibitem[{{Campbell} {et~al.}(2023){Campbell}, {Keys}, {Mathioudakis},
  {W{\"o}ger}, {Schad}, {Tritschler}, {de Wijn}, {Smitha}, {Beck}, {Christian},
  {Jess}, \& {Erd{\'e}lyi}}]{Campbell2023ApJ}
{Campbell}, R.~J., {Keys}, P.~H., {Mathioudakis}, M., {et~al.} 2023,
  \href{http://dx.doi.org/10.3847/2041-8213/acf85d}{\color{magenta}\apjl},
  \href{https://ui.adsabs.harvard.edu/abs/2023ApJ...955L..36C}{955, L36}

\bibitem[{{Castelli} \& {Kurucz}(2004)}]{Castelli2004A&A...419..725C}
{Castelli}, F. \& {Kurucz}, R.~L. 2004,
  \href{http://dx.doi.org/10.1051/0004-6361:20040079}{\color{magenta}\aap},
  \href{https://ui.adsabs.harvard.edu/abs/2004A&A...419..725C}{419, 725}

\bibitem[{{Centeno} {et~al.}(2022){Centeno}, {Flyer}, {Mukherjee}, {Egeland},
  {Casini}, {del Pino Alem{\'a}n}, \& {Rempel}}]{Centeno2022ApJ}
{Centeno}, R., {Flyer}, N., {Mukherjee}, L., {et~al.} 2022,
  \href{http://dx.doi.org/10.3847/1538-4357/ac402f}{\color{magenta}\apj},
  \href{https://ui.adsabs.harvard.edu/abs/2022ApJ...925..176C}{925, 176}

\bibitem[{{Chappell} \& {Pereira}(2022)}]{Chappell2022AA}
{Chappell}, B.~A. \& {Pereira}, T. M.~D. 2022,
  \href{http://dx.doi.org/10.1051/0004-6361/202142625}{\color{magenta}\aap},
  \href{https://ui.adsabs.harvard.edu/abs/2022A&A...658A.182C}{658, A182}

\bibitem[{{Chen} {et~al.}(2023{\natexlab{a}}){Chen}, {Cheung}, {Rempel}, \&
  {Chintzoglou}}]{ChenF2023ApJ}
{Chen}, F., {Cheung}, M. C.~M., {Rempel}, M., \& {Chintzoglou}, G.
  2023{\natexlab{a}},
  \href{http://dx.doi.org/10.3847/1538-4357/acc8c5}{\color{magenta}\apj},
  \href{https://ui.adsabs.harvard.edu/abs/2023ApJ...949..118C}{949, 118}

\bibitem[{{Chen} {et~al.}(2017){Chen}, {Rempel}, \& {Fan}}]{ChenF2017}
{Chen}, F., {Rempel}, M., \& {Fan}, Y. 2017,
  \href{http://dx.doi.org/10.3847/1538-4357/aa85a0}{\color{magenta}\apj},
  \href{https://ui.adsabs.harvard.edu/abs/2017ApJ...846..149C}{846, 149}

\bibitem[{{Chen} {et~al.}(2023{\natexlab{b}}){Chen}, {Rempel}, \&
  {Fan}}]{ChenF2023ApJL}
{Chen}, F., {Rempel}, M., \& {Fan}, Y. 2023{\natexlab{b}},
  \href{http://dx.doi.org/10.3847/2041-8213/acda2e}{\color{magenta}\apjl},
  \href{https://ui.adsabs.harvard.edu/abs/2023ApJ...950L...3C}{950, L3}

\bibitem[{{Cheung} {et~al.}(2019){Cheung}, {Rempel}, {Chintzoglou}, {Chen},
  {Testa}, {Mart{\'\i}nez-Sykora}, {Sainz Dalda}, {DeRosa}, {Malanushenko},
  {Hansteen}, {De Pontieu}, {Carlsson}, {Gudiksen}, \& {McIntosh}}]{cheung2019}
{Cheung}, M.~C.~M., {Rempel}, M., {Chintzoglou}, G., {et~al.} 2019,
  \href{http://dx.doi.org/10.1038/s41550-018-0629-3}{\color{magenta}Nat.
  Astron.}, \href{https://ui.adsabs.harvard.edu/abs/2019NatAs...3..160C}{3,
  160}

\bibitem[{{Cheung} {et~al.}(2010){Cheung}, {Rempel}, {Title}, \&
  {Sch{\"u}ssler}}]{cheung2010}
{Cheung}, M.~C.~M., {Rempel}, M., {Title}, A.~M., \& {Sch{\"u}ssler}, M. 2010,
  \href{http://dx.doi.org/10.1088/0004-637X/720/1/233}{\color{magenta}Astrophys.
  J.}, \href{https://ui.adsabs.harvard.edu/abs/2010ApJ...720..233C}{720, 233}

\bibitem[{{Chitta} {et~al.}(2017){Chitta}, {Peter}, {Young}, \&
  {Huang}}]{Chitta2017AA}
{Chitta}, L.~P., {Peter}, H., {Young}, P.~R., \& {Huang}, Y.~M. 2017,
  \href{http://dx.doi.org/10.1051/0004-6361/201730830}{\color{magenta}\aap},
  \href{https://ui.adsabs.harvard.edu/abs/2017A&A...605A..49C}{605, A49}

\bibitem[{{da Silva Santos} {et~al.}(2023){da Silva Santos}, {Reardon},
  {Cauzzi}, {Schad}, {Mart{\'\i}nez Pillet}, {Tritschler}, {W{\"o}ger},
  {Hofmann}, {Stauffer}, \& {Uitenbroek}}]{daSilvaSantos2023ApJ}
{da Silva Santos}, J.~M., {Reardon}, K., {Cauzzi}, G., {et~al.} 2023,
  \href{http://dx.doi.org/10.3847/2041-8213/acf21f}{\color{magenta}\apjl},
  \href{https://ui.adsabs.harvard.edu/abs/2023ApJ...954L..35D}{954, L35}

\bibitem[{{de Wijn} {et~al.}(2022){de Wijn}, {Casini}, {Carlile}, {Lecinski},
  {Sewell}, {Zmarzly}, {Eigenbrot}, {Beck}, {W{\"o}ger}, \&
  {Kn{\"o}lker}}]{deWijn2022SoPh}
{de Wijn}, A.~G., {Casini}, R., {Carlile}, A., {et~al.} 2022,
  \href{http://dx.doi.org/10.1007/s11207-022-01954-1}{\color{magenta}\solphys},
  \href{https://ui.adsabs.harvard.edu/abs/2022SoPh..297...22D}{297, 22}

\bibitem[{{del Toro Iniesta}(2007)}]{delToroIniesta2007book}
{del Toro Iniesta}, J.~C. 2007, {Introduction to Spectropolarimetry}

\bibitem[{del Toro~Iniesta \& Ruiz~Cobo(2016)}]{delToroIniesta2016}
del Toro~Iniesta, J.~C. \& Ruiz~Cobo, B. 2016,
  \href{http://dx.doi.org/10.1007/s41116-016-0005-2}{\color{magenta}Living
  Reviews in Solar Physics}, 13, 13

\bibitem[{{delaCruzRodriguez} {et~al.}(2019){delaCruzRodriguez}, {Leenaarts},
  {Danilovic}, \& {Uitenbroek}}]{delaCruzRodriguez2019AA}
{delaCruzRodriguez}, J., {Leenaarts}, J., {Danilovic}, S., \& {Uitenbroek}, H.
  2019,
  \href{http://dx.doi.org/10.1051/0004-6361/201834464}{\color{magenta}\aap},
  \href{https://ui.adsabs.harvard.edu/abs/2019A&A...623A..74D}{623, A74}

\bibitem[{{D{\'\i}az Baso} \& {Asensio
  Ramos}(2018)}]{DiazBaso2018A&A...614A...5D}
{D{\'\i}az Baso}, C.~J. \& {Asensio Ramos}, A. 2018,
  \href{http://dx.doi.org/10.1051/0004-6361/201731344}{\color{magenta}\aap},
  \href{https://ui.adsabs.harvard.edu/abs/2018A&A...614A...5D}{614, A5}

\bibitem[{{D{\'\i}az Baso} {et~al.}(2022){D{\'\i}az Baso}, {Asensio Ramos}, \&
  {de la Cruz Rodr{\'\i}guez}}]{DiazBaso2022AA}
{D{\'\i}az Baso}, C.~J., {Asensio Ramos}, A., \& {de la Cruz Rodr{\'\i}guez},
  J. 2022,
  \href{http://dx.doi.org/10.1051/0004-6361/202142018}{\color{magenta}\aap},
  \href{https://ui.adsabs.harvard.edu/abs/2022A&A...659A.165D}{659, A165}

\bibitem[{{D{\'\i}az Baso} {et~al.}(2019){D{\'\i}az Baso}, {de la Cruz
  Rodr{\'\i}guez}, \& {Danilovic}}]{DiazBaso2019A&A...629A..99D}
{D{\'\i}az Baso}, C.~J., {de la Cruz Rodr{\'\i}guez}, J., \& {Danilovic}, S.
  2019,
  \href{http://dx.doi.org/10.1051/0004-6361/201936069}{\color{magenta}\aap},
  \href{https://ui.adsabs.harvard.edu/abs/2019A&A...629A..99D}{629, A99}

\bibitem[{{Eklund}(2023)}]{Eklund2023A&A...669A.106E}
{Eklund}, H. 2023,
  \href{http://dx.doi.org/10.1051/0004-6361/202244484}{\color{magenta}\aap},
  \href{https://ui.adsabs.harvard.edu/abs/2023A&A...669A.106E}{669, A106}

\bibitem[{{Florios} {et~al.}(2018){Florios}, {Kontogiannis}, {Park}, {Guerra},
  {Benvenuto}, {Bloomfield}, \& {Georgoulis}}]{Florios2018SoPh}
{Florios}, K., {Kontogiannis}, I., {Park}, S.-H., {et~al.} 2018,
  \href{http://dx.doi.org/10.1007/s11207-018-1250-4}{\color{magenta}\solphys},
  \href{https://ui.adsabs.harvard.edu/abs/2018SoPh..293...28F}{293, 28}

\bibitem[{{Gafeira} {et~al.}(2021){Gafeira}, {Orozco Su{\'a}rez}, {Mili{\'c}},
  {Quintero Noda}, {Ruiz Cobo}, \& {Uitenbroek}}]{Gafeira2021AA}
{Gafeira}, R., {Orozco Su{\'a}rez}, D., {Mili{\'c}}, I., {et~al.} 2021,
  \href{http://dx.doi.org/10.1051/0004-6361/201936910}{\color{magenta}\aap},
  \href{https://ui.adsabs.harvard.edu/abs/2021A&A...651A..31G}{651, A31}

\bibitem[{{Goodwin} {et~al.}(2024){Goodwin}, {Sadykov}, \&
  {Martens}}]{Goodwin2024arXiv}
{Goodwin}, G.~T., {Sadykov}, V.~M., \& {Martens}, P.~C. 2024,
  \href{https://ui.adsabs.harvard.edu/abs/2024arXiv240205288G}{\href{http://dx.doi.org/10.48550/arXiv.2402.05288}{\color{magenta}arXiv
  e-prints}, arXiv:2402.05288}

\bibitem[{{Grevesse} \& {Sauval}(1998)}]{Grevesse1998SSRv}
{Grevesse}, N. \& {Sauval}, A.~J. 1998,
  \href{http://dx.doi.org/10.1023/A:1005161325181}{\color{magenta}\ssr},
  \href{https://ui.adsabs.harvard.edu/abs/1998SSRv...85..161G}{85, 161}

\bibitem[{{Harrington} {et~al.}(2023){Harrington}, {Sueoka}, {Schad}, {Beck},
  {Eigenbrot}, {de Wijn}, {Casini}, {White}, \& {Jaeggli}}]{Harrington2023SoPh}
{Harrington}, D.~M., {Sueoka}, S.~R., {Schad}, T.~A., {et~al.} 2023,
  \href{http://dx.doi.org/10.1007/s11207-022-02101-6}{\color{magenta}\solphys},
  \href{https://ui.adsabs.harvard.edu/abs/2023SoPh..298...10H}{298, 10}

\bibitem[{{Higgins} {et~al.}(2022){Higgins}, {Fouhey}, {Antiochos}, {Barnes},
  {Cheung}, {Hoeksema}, {Leka}, {Liu}, {Schuck}, \&
  {Gombosi}}]{Higgins2022ApJS}
{Higgins}, R. E.~L., {Fouhey}, D.~F., {Antiochos}, S.~K., {et~al.} 2022,
  \href{http://dx.doi.org/10.3847/1538-4365/ac42d5}{\color{magenta}\apjs},
  \href{https://ui.adsabs.harvard.edu/abs/2022ApJS..259...24H}{259, 24}

\bibitem[{{Higgins} {et~al.}(2021){Higgins}, {Fouhey}, {Zhang}, {Antiochos},
  {Barnes}, {Hoeksema}, {Leka}, {Liu}, {Schuck}, \& {Gombosi}}]{Higgins2021ApJ}
{Higgins}, R. E.~L., {Fouhey}, D.~F., {Zhang}, D., {et~al.} 2021,
  \href{http://dx.doi.org/10.3847/1538-4357/abd7fe}{\color{magenta}\apj},
  \href{https://ui.adsabs.harvard.edu/abs/2021ApJ...911..130H}{911, 130}

\bibitem[{{Huang} {et~al.}(2018){Huang}, {Wang}, {Xu}, {Liu}, {Li}, \&
  {Dai}}]{Huang2018ApJ}
{Huang}, X., {Wang}, H., {Xu}, L., {et~al.} 2018,
  \href{http://dx.doi.org/10.3847/1538-4357/aaae00}{\color{magenta}\apj},
  \href{https://ui.adsabs.harvard.edu/abs/2018ApJ...856....7H}{856, 7}

\bibitem[{{Hubeny} \& {Mihalas}(2015)}]{Hubeny2015book}
{Hubeny}, I. \& {Mihalas}, D. 2015, {Theory of Stellar Atmospheres. An
  Introduction to Astrophysical Non-equilibrium Quantitative Spectroscopic
  Analysis}

\bibitem[{Hunter(2007)}]{Hunter2007}
Hunter, J.~D. 2007,
  \href{http://dx.doi.org/10.1109/MCSE.2007.55}{\color{magenta}Computing in
  Science \& Engineering}, 9, 9

\bibitem[{{Jaeggli} {et~al.}(2022){Jaeggli}, {Lin}, {Onaka}, {Yamada}, {Anan},
  {Bonnet}, {Ching}, {Huang}, {Kramar}, {McGregor}, {Nitta}, {Rae},
  {Robertson}, {Schad}, {Toyama}, {Young}, {Berst}, {Harrington}, {Liang},
  {Puentes}, {Sekulic}, {Smith}, \& {Sueoka}}]{Jaeggli2022SoPh}
{Jaeggli}, S.~A., {Lin}, H., {Onaka}, P., {et~al.} 2022,
  \href{http://dx.doi.org/10.1007/s11207-022-02062-w}{\color{magenta}\solphys},
  \href{https://ui.adsabs.harvard.edu/abs/2022SoPh..297..137J}{297, 137}

\bibitem[{{Jarolim} {et~al.}(2023){Jarolim}, {Thalmann}, {Veronig}, \&
  {Podladchikova}}]{Jarolim2023NatAs}
{Jarolim}, R., {Thalmann}, J.~K., {Veronig}, A.~M., \& {Podladchikova}, T.
  2023,
  \href{http://dx.doi.org/10.1038/s41550-023-02030-9}{\color{magenta}Nature
  Astronomy}, \href{https://ui.adsabs.harvard.edu/abs/2023NatAs...7.1171J}{7,
  1171}

\bibitem[{{Jarolim} {et~al.}(2024{\natexlab{a}}){Jarolim}, {Tremblay},
  {Mu{\~n}oz-Jaramillo}, {Bintsi}, {Jungbluth}, {Santos}, {Vourlidas}, {Mason},
  {Sundaresan}, {Downs}, \& {Caplan}}]{Jarolim2024ApJb}
{Jarolim}, R., {Tremblay}, B., {Mu{\~n}oz-Jaramillo}, A., {et~al.}
  2024{\natexlab{a}},
  \href{http://dx.doi.org/10.3847/2041-8213/ad12d2}{\color{magenta}\apjl},
  \href{https://ui.adsabs.harvard.edu/abs/2024ApJ...961L..31J}{961, L31}

\bibitem[{{Jarolim} {et~al.}(2024{\natexlab{b}}){Jarolim}, {Tremblay},
  {Rempel}, {Molnar}, {Veronig}, {Thalmann}, \&
  {Podladchikova}}]{Jarolim2024ApJa}
{Jarolim}, R., {Tremblay}, B., {Rempel}, M., {et~al.} 2024{\natexlab{b}},
  \href{http://dx.doi.org/10.3847/2041-8213/ad2450}{\color{magenta}\apjl},
  \href{https://ui.adsabs.harvard.edu/abs/2024ApJ...963L..21J}{963, L21}

\bibitem[{{Kazachenko} {et~al.}(2014){Kazachenko}, {Fisher}, \&
  {Welsch}}]{kazachenko2014}
{Kazachenko}, M.~D., {Fisher}, G.~H., \& {Welsch}, B.~T. 2014,
  \href{http://dx.doi.org/10.1088/0004-637X/795/1/17}{\color{magenta}\apj},
  \href{https://ui.adsabs.harvard.edu/abs/2014ApJ...795...17K}{795, 17}

\bibitem[{{Kuridze} {et~al.}(2024){Kuridze}, {Uitenbroek}, {W{\"o}ger},
  {Mathioudakis}, {Morgan}, {Campbell}, {Fischer}, {Cauzzi}, {Schad},
  {Reardon}, {da Silva Santos}, {Beck}, {Tritschler}, \&
  {Rimmele}}]{Kuridze2024}
{Kuridze}, D., {Uitenbroek}, H., {W{\"o}ger}, F., {et~al.} 2024,
  \href{https://ui.adsabs.harvard.edu/abs/2024arXiv240204545K}{\href{http://dx.doi.org/10.48550/arXiv.2402.04545}{\color{magenta}arXiv
  e-prints}, arXiv:2402.04545}

\bibitem[{{Lites} {et~al.}(2008){Lites}, {Kubo}, {Socas-Navarro}, {Berger},
  {Frank}, {Shine}, {Tarbell}, {Title}, {Ichimoto}, {Katsukawa}, {Tsuneta},
  {Suematsu}, {Shimizu}, \& {Nagata}}]{lites2008}
{Lites}, B.~W., {Kubo}, M., {Socas-Navarro}, H., {et~al.} 2008,
  \href{http://dx.doi.org/10.1086/522922}{\color{magenta}\apj},
  \href{https://ui.adsabs.harvard.edu/abs/2008ApJ...672.1237L}{672, 1237}

\bibitem[{{Liu} {et~al.}(2016){Liu}, {Xu}, {Cao}, {Deng}, {Lee}, {Hudson},
  {Gary}, {Wang}, {Jing}, \& {Wang}}]{Liu2016NatCo}
{Liu}, C., {Xu}, Y., {Cao}, W., {et~al.} 2016,
  \href{http://dx.doi.org/10.1038/ncomms13104}{\color{magenta}Nature
  Communications},
  \href{https://ui.adsabs.harvard.edu/abs/2016NatCo...713104L}{7, 13104}

\bibitem[{{Liu} {et~al.}(2014){Liu}, {Hoeksema}, {Bobra}, {Hayashi}, {Schuck},
  \& {Sun}}]{Liu2014ApJ}
{Liu}, Y., {Hoeksema}, J.~T., {Bobra}, M., {et~al.} 2014,
  \href{http://dx.doi.org/10.1088/0004-637X/785/1/13}{\color{magenta}\apj},
  \href{https://ui.adsabs.harvard.edu/abs/2014ApJ...785...13L}{785, 13}

\bibitem[{{Liu} \& {Schuck}(2012{\natexlab{a}})}]{Liu2012ApJ}
{Liu}, Y. \& {Schuck}, P.~W. 2012{\natexlab{a}},
  \href{http://dx.doi.org/10.1088/0004-637X/761/2/105}{\color{magenta}\apj},
  \href{https://ui.adsabs.harvard.edu/abs/2012ApJ...761..105L}{761, 105}

\bibitem[{{Liu} \& {Schuck}(2012{\natexlab{b}})}]{liuy2012}
{Liu}, Y. \& {Schuck}, P.~W. 2012{\natexlab{b}},
  \href{http://dx.doi.org/10.1088/0004-637X/761/2/105}{\color{magenta}\apj},
  \href{https://ui.adsabs.harvard.edu/abs/2012ApJ...761..105L}{761, 105}

\bibitem[{{Liu} {et~al.}(2023){Liu}, {Welsch}, {Valori}, {Georgoulis}, {Guo},
  {Pariat}, {Park}, \& {Thalmann}}]{Liu2023ApJ}
{Liu}, Y., {Welsch}, B.~T., {Valori}, G., {et~al.} 2023,
  \href{http://dx.doi.org/10.3847/1538-4357/aca3a6}{\color{magenta}\apj},
  \href{https://ui.adsabs.harvard.edu/abs/2023ApJ...942...27L}{942, 27}

\bibitem[{{Lord}(2014)}]{lord2014b}
{Lord}, J.~W. 2014,
  \href{https://ui.adsabs.harvard.edu/abs/2014PhDT.......241L}{{Deep
  Convection, Magnetism and Solar Supergranulation}}, PhD thesis, University of
  Colorado, Boulder

\bibitem[{{Lord} {et~al.}(2014){Lord}, {Cameron}, {Rast}, {Rempel}, \&
  {Roudier}}]{Lord2014ApJ}
{Lord}, J.~W., {Cameron}, R.~H., {Rast}, M.~P., {Rempel}, M., \& {Roudier}, T.
  2014,
  \href{http://dx.doi.org/10.1088/0004-637X/793/1/24}{\color{magenta}\apj},
  \href{https://ui.adsabs.harvard.edu/abs/2014ApJ...793...24L}{793, 24}

\bibitem[{{Lumme} {et~al.}(2019){Lumme}, {Kazachenko}, {Fisher}, {Welsch},
  {Pomoell}, \& {Kilpua}}]{Lumme2019SoPh}
{Lumme}, E., {Kazachenko}, M.~D., {Fisher}, G.~H., {et~al.} 2019,
  \href{http://dx.doi.org/10.1007/s11207-019-1475-x}{\color{magenta}\solphys},
  \href{https://ui.adsabs.harvard.edu/abs/2019SoPh..294...84L}{294, 84}

\bibitem[{{Mili{\'c}} \& {Gafeira}(2020)}]{Milic2020AA}
{Mili{\'c}}, I. \& {Gafeira}, R. 2020,
  \href{http://dx.doi.org/10.1051/0004-6361/201936537}{\color{magenta}\aap},
  \href{https://ui.adsabs.harvard.edu/abs/2020A&A...644A.129M}{644, A129}

\bibitem[{{Mili{\'c}} \& {van Noort}(2018)}]{Milic2018AA}
{Mili{\'c}}, I. \& {van Noort}, M. 2018,
  \href{http://dx.doi.org/10.1051/0004-6361/201833382}{\color{magenta}\aap},
  \href{https://ui.adsabs.harvard.edu/abs/2018A&A...617A..24M}{617, A24}

\bibitem[{{Mistryukova} {et~al.}(2023){Mistryukova}, {Plotnikov}, {Khizhik},
  {Knyazeva}, {Hushchyn}, \& {Derkach}}]{Mistryukova2023SoPh}
{Mistryukova}, L., {Plotnikov}, A., {Khizhik}, A., {et~al.} 2023,
  \href{http://dx.doi.org/10.1007/s11207-023-02189-4}{\color{magenta}\solphys},
  \href{https://ui.adsabs.harvard.edu/abs/2023SoPh..298...98M}{298, 98}

\bibitem[{{Moore} {et~al.}(2001){Moore}, {Sterling}, {Hudson}, \&
  {Lemen}}]{Moore2001ApJ}
{Moore}, R.~L., {Sterling}, A.~C., {Hudson}, H.~S., \& {Lemen}, J.~R. 2001,
  \href{http://dx.doi.org/10.1086/320559}{\color{magenta}\apj},
  \href{https://ui.adsabs.harvard.edu/abs/2001ApJ...552..833M}{552, 833}

\bibitem[{{Nishizuka} {et~al.}(2018){Nishizuka}, {Sugiura}, {Kubo}, {Den}, \&
  {Ishii}}]{Nishizuka2018ApJ}
{Nishizuka}, N., {Sugiura}, K., {Kubo}, Y., {Den}, M., \& {Ishii}, M. 2018,
  \href{http://dx.doi.org/10.3847/1538-4357/aab9a7}{\color{magenta}\apj},
  \href{https://ui.adsabs.harvard.edu/abs/2018ApJ...858..113N}{858, 113}

\bibitem[{{Pastor Yabar} {et~al.}(2019){Pastor Yabar}, {Borrero}, \& {Ruiz
  Cobo}}]{PastorYabar2019AA}
{Pastor Yabar}, A., {Borrero}, J.~M., \& {Ruiz Cobo}, B. 2019,
  \href{http://dx.doi.org/10.1051/0004-6361/201935692}{\color{magenta}\aap},
  \href{https://ui.adsabs.harvard.edu/abs/2019A&A...629A..24P}{629, A24}

\bibitem[{{Priest}(2014)}]{Pries2014book}
{Priest}, E. 2014, {Magnetohydrodynamics of the Sun}

\bibitem[{{Priest} \& {Forbes}(2002)}]{Priest2002AARv}
{Priest}, E.~R. \& {Forbes}, T.~G. 2002,
  \href{http://dx.doi.org/10.1007/s001590100013}{\color{magenta}\aapr},
  \href{https://ui.adsabs.harvard.edu/abs/2002A&ARv..10..313P}{10, 313}

\bibitem[{{Quintero Noda} {et~al.}(2021){Quintero Noda}, {Barklem}, {Gafeira},
  {Ruiz Cobo}, {Collados}, {Carlsson}, {Mart{\'\i}nez Pillet}, {Orozco
  Su{\'a}rez}, {Uitenbroek}, \& {Katsukawa}}]{QuinteroNoda2021AA}
{Quintero Noda}, C., {Barklem}, P.~S., {Gafeira}, R., {et~al.} 2021,
  \href{http://dx.doi.org/10.1051/0004-6361/202037735}{\color{magenta}\aap},
  \href{https://ui.adsabs.harvard.edu/abs/2021A&A...652A.161Q}{652, A161}

\bibitem[{{Quintero Noda} {et~al.}(2023){Quintero Noda}, {Khomenko},
  {Collados}, {Ruiz Cobo}, {Gafeira}, {Vitas}, {Rempel}, {Campbell}, {Pastor
  Yabar}, {Uitenbroek}, \& {Orozco Su{\'a}rez}}]{QuinteroNoda2023AA}
{Quintero Noda}, C., {Khomenko}, E., {Collados}, M., {et~al.} 2023,
  \href{http://dx.doi.org/10.1051/0004-6361/202345890}{\color{magenta}\aap},
  \href{https://ui.adsabs.harvard.edu/abs/2023A&A...675A..93Q}{675, A93}

\bibitem[{{Rahman} {et~al.}(2024){Rahman}, {Jeong}, {Siddique}, {Moon}, \&
  {Lawrance}}]{Rahman2024ApJS}
{Rahman}, S., {Jeong}, H.-J., {Siddique}, A., {Moon}, Y.-J., \& {Lawrance}, B.
  2024,
  \href{http://dx.doi.org/10.3847/1538-4365/ad1877}{\color{magenta}\apjs},
  \href{https://ui.adsabs.harvard.edu/abs/2024ApJS..271...14R}{271, 14}

\bibitem[{{Rahman} {et~al.}(2020){Rahman}, {Moon}, {Park}, {Siddique}, {Cho},
  \& {Lim}}]{Rahman2020ApJ...897L..32R}
{Rahman}, S., {Moon}, Y.-J., {Park}, E., {et~al.} 2020,
  \href{http://dx.doi.org/10.3847/2041-8213/ab9d79}{\color{magenta}\apjl},
  \href{https://ui.adsabs.harvard.edu/abs/2020ApJ...897L..32R}{897, L32}

\bibitem[{{Rahman} {et~al.}(2023){Rahman}, {Shin}, {Jeong}, {Siddique}, {Moon},
  {Park}, {Kang}, \& {Bae}}]{Rahman2023ApJ}
{Rahman}, S., {Shin}, S., {Jeong}, H.-J., {et~al.} 2023,
  \href{http://dx.doi.org/10.3847/1538-4357/acbd3c}{\color{magenta}\apj},
  \href{https://ui.adsabs.harvard.edu/abs/2023ApJ...948...21R}{948, 21}

\bibitem[{{Rast} {et~al.}(2021){Rast}, {Bello Gonz{\'a}lez}, {Bellot Rubio},
  {Cao}, {Cauzzi}, {Deluca}, {de Pontieu}, {Fletcher}, {Gibson}, {Judge},
  {Katsukawa}, {Kazachenko}, {Khomenko}, {Landi}, {Mart{\'\i}nez Pillet},
  {Petrie}, {Qiu}, {Rachmeler}, {Rempel}, {Schmidt}, {Scullion}, {Sun},
  {Welsch}, {Andretta}, {Antolin}, {Ayres}, {Balasubramaniam}, {Ballai},
  {Berger}, {Bradshaw}, {Campbell}, {Carlsson}, {Casini}, {Centeno}, {Cranmer},
  {Criscuoli}, {Deforest}, {Deng}, {Erd{\'e}lyi}, {Fedun}, {Fischer},
  {Gonz{\'a}lez Manrique}, {Hahn}, {Harra}, {Henriques}, {Hurlburt}, {Jaeggli},
  {Jafarzadeh}, {Jain}, {Jefferies}, {Keys}, {Kowalski}, {Kuckein}, {Kuhn},
  {Kuridze}, {Liu}, {Liu}, {Longcope}, {Mathioudakis}, {McAteer}, {McIntosh},
  {McKenzie}, {Miralles}, {Morton}, {Muglach}, {Nelson}, {Panesar}, {Parenti},
  {Parnell}, {Poduval}, {Reardon}, {Reep}, {Schad}, {Schmit}, {Sharma},
  {Socas-Navarro}, {Srivastava}, {Sterling}, {Suematsu}, {Tarr}, {Tiwari},
  {Tritschler}, {Verth}, {Vourlidas}, {Wang}, {Wang}, {NSO and DKIST Project},
  {DKIST Instrument Scientists}, {DKIST Science Working Group}, \& {DKIST
  Critical Science Plan Community}}]{rast2021}
{Rast}, M.~P., {Bello Gonz{\'a}lez}, N., {Bellot Rubio}, L., {et~al.} 2021,
  \href{http://dx.doi.org/10.1007/s11207-021-01789-2}{\color{magenta}\solphys},
  \href{https://ui.adsabs.harvard.edu/abs/2021SoPh..296...70R}{296, 70}

\bibitem[{{Rempel}(2012)}]{rempel2012}
{Rempel}, M. 2012,
  \href{http://dx.doi.org/10.1088/0004-637X/750/1/62}{\color{magenta}Astrophys.
  J.}, \href{https://ui.adsabs.harvard.edu/abs/2012ApJ...750...62R}{750, 62}

\bibitem[{{Rempel}(2014)}]{rempel2014}
{Rempel}, M. 2014,
  \href{http://dx.doi.org/10.1088/0004-637X/789/2/132}{\color{magenta}\apj},
  \href{https://ui.adsabs.harvard.edu/abs/2014ApJ...789..132R}{789, 132}

\bibitem[{{Rempel}(2017)}]{rempel2017}
{Rempel}, M. 2017,
  \href{http://dx.doi.org/10.3847/1538-4357/834/1/10}{\color{magenta}\apj},
  \href{https://ui.adsabs.harvard.edu/abs/2017ApJ...834...10R}{834, 10}

\bibitem[{{Rempel} {et~al.}(2009){Rempel}, {Sch{\"u}ssler}, \&
  {Kn{\"o}lker}}]{Rempel2009ApJ}
{Rempel}, M., {Sch{\"u}ssler}, M., \& {Kn{\"o}lker}, M. 2009,
  \href{http://dx.doi.org/10.1088/0004-637X/691/1/640}{\color{magenta}\apj},
  \href{https://ui.adsabs.harvard.edu/abs/2009ApJ...691..640R}{691, 640}

\bibitem[{{Rimmele} {et~al.}(2020){Rimmele}, {Warner}, {Keil}, {Goode},
  {Kn{\"o}lker}, {Kuhn}, {Rosner}, {McMullin}, {Casini}, {Lin}, {W{\"o}ger},
  {von der L{\"u}he}, {Tritschler}, {Davey}, {de Wijn}, {Elmore}, {Fehlmann},
  {Harrington}, {Jaeggli}, {Rast}, {Schad}, {Schmidt}, {Mathioudakis},
  {Mickey}, {Anan}, {Beck}, {Marshall}, {Jeffers}, {Oschmann}, {Beard},
  {Berst}, {Cowan}, {Craig}, {Cross}, {Cummings}, {Donnelly}, {de Vanssay},
  {Eigenbrot}, {Ferayorni}, {Foster}, {Galapon}, {Gedrites}, {Gonzales},
  {Goodrich}, {Gregory}, {Guzman}, {Guzzo}, {Hegwer}, {Hubbard}, {Hubbard},
  {Johansson}, {Johnson}, {Liang}, {Liang}, {McQuillen}, {Mayer}, {Newman},
  {Onodera}, {Phelps}, {Puentes}, {Richards}, {Rimmele}, {Sekulic}, {Shimko},
  {Simison}, {Smith}, {Starman}, {Sueoka}, {Summers}, {Szabo}, {Szabo},
  {Wampler}, {Williams}, \& {White}}]{Rimmele2020SoPh}
{Rimmele}, T.~R., {Warner}, M., {Keil}, S.~L., {et~al.} 2020,
  \href{http://dx.doi.org/10.1007/s11207-020-01736-7}{\color{magenta}\solphys},
  \href{https://ui.adsabs.harvard.edu/abs/2020SoPh..295..172R}{295, 172}

\bibitem[{{Ruiz Cobo} \& {del Toro Iniesta}(1992)}]{RuizCobo1992}
{Ruiz Cobo}, B. \& {del Toro Iniesta}, J.~C. 1992,
  \href{http://dx.doi.org/10.1086/171862}{\color{magenta}\apj},
  \href{https://ui.adsabs.harvard.edu/abs/1992ApJ...398..375R}{398, 375}

\bibitem[{{Ruiz Cobo} \& {del Toro Iniesta}(2012)}]{ruizcobo2012}
{Ruiz Cobo}, B. \& {del Toro Iniesta}, J.~C. 2012, {SIR: Stokes Inversion based
  on Response functions}, Astrophysics Source Code Library

\bibitem[{{Ruiz Cobo} {et~al.}(2022){Ruiz Cobo}, {Quintero Noda}, {Gafeira},
  {Uitenbroek}, {Orozco Su{\'a}rez}, \& {P{\'a}ez Ma{\~n}{\'a}}}]{RuizCobo2022}
{Ruiz Cobo}, B., {Quintero Noda}, C., {Gafeira}, R., {et~al.} 2022,
  \href{http://dx.doi.org/10.1051/0004-6361/202140877}{\color{magenta}\aap},
  \href{https://ui.adsabs.harvard.edu/abs/2022A&A...660A..37R}{660, A37}

\bibitem[{Sadowski \& Baldi(2018)}]{Sadowski2018}
Sadowski, P. \& Baldi, P. 2018, Deep Learning in the Natural Sciences:
  Applications to Physics, ed. L.~Rozonoer, B.~Mirkin, \& I.~Muchnik (Cham:
  Springer International Publishing), 269--297

\bibitem[{{Sainz Dalda} {et~al.}(2019){Sainz Dalda}, {de la Cruz
  Rodr{\'\i}guez}, {De Pontieu}, \& {Go{\v{s}}i{\'c}}}]{SainzDalda2019ApJ}
{Sainz Dalda}, A., {de la Cruz Rodr{\'\i}guez}, J., {De Pontieu}, B., \&
  {Go{\v{s}}i{\'c}}, M. 2019,
  \href{http://dx.doi.org/10.3847/2041-8213/ab15d9}{\color{magenta}\apjl},
  \href{https://ui.adsabs.harvard.edu/abs/2019ApJ...875L..18S}{875, L18}

\bibitem[{{Samanta} {et~al.}(2019){Samanta}, {Tian}, {Yurchyshyn}, {Peter},
  {Cao}, {Sterling}, {Erd{\'e}lyi}, {Ahn}, {Feng}, {Utz}, {Banerjee}, \&
  {Chen}}]{Samanta2019Sci}
{Samanta}, T., {Tian}, H., {Yurchyshyn}, V., {et~al.} 2019,
  \href{http://dx.doi.org/10.1126/science.aaw2796}{\color{magenta}Science},
  \href{https://ui.adsabs.harvard.edu/abs/2019Sci...366..890S}{366, 890}

\bibitem[{{Schad} {et~al.}(2023){Schad}, {Kuhn}, {Fehlmann}, {Scholl},
  {Harrington}, {Rimmele}, \& {Tritschler}}]{Schad2023ApJ}
{Schad}, T.~A., {Kuhn}, J.~R., {Fehlmann}, A., {et~al.} 2023,
  \href{http://dx.doi.org/10.3847/1538-4357/acabbd}{\color{magenta}\apj},
  \href{https://ui.adsabs.harvard.edu/abs/2023ApJ...943...59S}{943, 59}

\bibitem[{{Schmidt} {et~al.}(2016){Schmidt}, {Schubert}, {Ellwarth},
  {Baumgartner}, {Bell}, {Fischer}, {Halbgewachs}, {Heidecke}, {Kentischer},
  {von der L{\"u}he}, {Scheiffelen}, \& {Sigwarth}}]{Schmidt2016SPIE}
{Schmidt}, W., {Schubert}, M., {Ellwarth}, M., {et~al.} 2016, in Society of
  Photo-Optical Instrumentation Engineers (SPIE) Conference Series, Vol. 9908,
  Ground-based and Airborne Instrumentation for Astronomy VI, ed. C.~J.
  {Evans}, L.~{Simard}, \& H.~{Takami},
  \href{https://ui.adsabs.harvard.edu/abs/2016SPIE.9908E..4NS}{99084N}

\bibitem[{{Schuck} \& {Antiochos}(2019)}]{schuck2019}
{Schuck}, P.~W. \& {Antiochos}, S.~K. 2019,
  \href{http://dx.doi.org/10.3847/1538-4357/ab298a}{\color{magenta}\apj},
  \href{https://ui.adsabs.harvard.edu/abs/2019ApJ...882..151S}{882, 151}

\bibitem[{{Smitha} {et~al.}(2020){Smitha}, {Holzreuter}, {van Noort}, \&
  {Solanki}}]{Smitha2020AA}
{Smitha}, H.~N., {Holzreuter}, R., {van Noort}, M., \& {Solanki}, S.~K. 2020,
  \href{http://dx.doi.org/10.1051/0004-6361/201937041}{\color{magenta}\aap},
  \href{https://ui.adsabs.harvard.edu/abs/2020A&A...633A.157S}{633, A157}

\bibitem[{{Smitha} {et~al.}(2021){Smitha}, {Holzreuter}, {van Noort}, \&
  {Solanki}}]{Smitha2021AA}
{Smitha}, H.~N., {Holzreuter}, R., {van Noort}, M., \& {Solanki}, S.~K. 2021,
  \href{http://dx.doi.org/10.1051/0004-6361/202039107}{\color{magenta}\aap},
  \href{https://ui.adsabs.harvard.edu/abs/2021A&A...647A..46S}{647, A46}

\bibitem[{{Socas-Navarro} {et~al.}(2015){Socas-Navarro}, {de la Cruz
  Rodr{\'\i}guez}, {Asensio Ramos}, {Trujillo Bueno}, \& {Ruiz
  Cobo}}]{SocasNavarro2015AA}
{Socas-Navarro}, H., {de la Cruz Rodr{\'\i}guez}, J., {Asensio Ramos}, A.,
  {Trujillo Bueno}, J., \& {Ruiz Cobo}, B. 2015,
  \href{http://dx.doi.org/10.1051/0004-6361/201424860}{\color{magenta}\aap},
  \href{https://ui.adsabs.harvard.edu/abs/2015A&A...577A...7S}{577, A7}

\bibitem[{{Song} {et~al.}(2022){Song}, {Ma}, {Ma}, {Zhao}, \&
  {Lin}}]{Song2022ApJS..263...25S}
{Song}, W., {Ma}, W., {Ma}, Y., {Zhao}, X., \& {Lin}, G. 2022,
  \href{http://dx.doi.org/10.3847/1538-4365/ac9a4d}{\color{magenta}\apjs},
  \href{https://ui.adsabs.harvard.edu/abs/2022ApJS..263...25S}{263, 25}

\bibitem[{{Stenflo}(1994)}]{Stenflo1994ASSL}
{Stenflo}, J. 1994, {Solar Magnetic Fields: Polarized Radiation Diagnostics},
  Vol. 189

\bibitem[{{Stenflo}(2012)}]{Stenflo2012AA}
{Stenflo}, J.~O. 2012,
  \href{http://dx.doi.org/10.1051/0004-6361/201219833}{\color{magenta}\aap},
  \href{https://ui.adsabs.harvard.edu/abs/2012A&A...547A..93S}{547, A93}

\bibitem[{{Sun} {et~al.}(2013){Sun}, {Hoeksema}, {Liu}, {Aulanier}, {Su},
  {Hannah}, \& {Hock}}]{Sun2013ApJ}
{Sun}, X., {Hoeksema}, J.~T., {Liu}, Y., {et~al.} 2013,
  \href{http://dx.doi.org/10.1088/0004-637X/778/2/139}{\color{magenta}\apj},
  \href{https://ui.adsabs.harvard.edu/abs/2013ApJ...778..139S}{778, 139}

\bibitem[{{Thalmann} {et~al.}(2021){Thalmann}, {Georgoulis}, {Liu}, {Pariat},
  {Valori}, {Anfinogentov}, {Chen}, {Guo}, {Moraitis}, {Yang}, {Mastrano}, \&
  {ISSI Team on Magnetic Helicity}}]{Thalmann2021ApJ}
{Thalmann}, J.~K., {Georgoulis}, M.~K., {Liu}, Y., {et~al.} 2021,
  \href{http://dx.doi.org/10.3847/1538-4357/ac1f93}{\color{magenta}\apj},
  \href{https://ui.adsabs.harvard.edu/abs/2021ApJ...922...41T}{922, 41}

\bibitem[{{Tremblay} {et~al.}(2018){Tremblay}, {Roudier}, {Rieutord}, \&
  {Vincent}}]{Tremblay2018SoPh..293...57T}
{Tremblay}, B., {Roudier}, T., {Rieutord}, M., \& {Vincent}, A. 2018,
  \href{http://dx.doi.org/10.1007/s11207-018-1276-7}{\color{magenta}\solphys},
  \href{https://ui.adsabs.harvard.edu/abs/2018SoPh..293...57T}{293, 57}

\bibitem[{{Uitenbroek}(2001)}]{Uitenbroek2001ApJ...557..389U}
{Uitenbroek}, H. 2001,
  \href{http://dx.doi.org/10.1086/321659}{\color{magenta}\apj},
  \href{https://ui.adsabs.harvard.edu/abs/2001ApJ...557..389U}{557, 389}

\bibitem[{{Vicente Ar{\'e}valo} {et~al.}(2022){Vicente Ar{\'e}valo}, {Asensio
  Ramos}, \& {Esteban Pozuelo}}]{VicenteArevalo2022ApJ}
{Vicente Ar{\'e}valo}, A., {Asensio Ramos}, A., \& {Esteban Pozuelo}, S. 2022,
  \href{http://dx.doi.org/10.3847/1538-4357/ac53b3}{\color{magenta}\apj},
  \href{https://ui.adsabs.harvard.edu/abs/2022ApJ...928..101V}{928, 101}

\bibitem[{Virtanen {et~al.}(2020)Virtanen, Gommers, Oliphant, Haberland, Reddy,
  Cournapeau, Burovski, Peterson, Weckesser, Bright, {van der Walt}, Brett,
  Wilson, Millman, Mayorov, Nelson, Jones, Kern, Larson, Carey, Polat, Feng,
  Moore, {VanderPlas}, Laxalde, Perktold, Cimrman, Henriksen, Quintero, Harris,
  Archibald, Ribeiro, Pedregosa, {van Mulbregt}, \& {SciPy 1.0
  Contributors}}]{2020SciPy-NMeth}
Virtanen, P., Gommers, R., Oliphant, T.~E., {et~al.} 2020,
  \href{http://dx.doi.org/10.1038/s41592-019-0686-2}{\color{magenta}Nature
  Methods}, \href{https://rdcu.be/b08Wh}{17, 261}

\bibitem[{{V{\"o}gler}(2004)}]{Vogler2004A&A...421..755V}
{V{\"o}gler}, A. 2004,
  \href{http://dx.doi.org/10.1051/0004-6361:20047044}{\color{magenta}\aap},
  \href{https://ui.adsabs.harvard.edu/abs/2004A&A...421..755V}{421, 755}

\bibitem[{{V{\"o}gler} \& {Sch{\"u}ssler}(2007)}]{Vogler2007AA}
{V{\"o}gler}, A. \& {Sch{\"u}ssler}, M. 2007,
  \href{http://dx.doi.org/10.1051/0004-6361:20077253}{\color{magenta}\aap},
  \href{https://ui.adsabs.harvard.edu/abs/2007A&A...465L..43V}{465, L43}

\bibitem[{{V{\"o}gler} {et~al.}(2005){V{\"o}gler}, {Shelyag}, {Sch{\"u}ssler},
  {Cattaneo}, {Emonet}, \& {Linde}}]{Vogler2005AA}
{V{\"o}gler}, A., {Shelyag}, S., {Sch{\"u}ssler}, M., {et~al.} 2005,
  \href{http://dx.doi.org/10.1051/0004-6361:20041507}{\color{magenta}\aap},
  \href{https://ui.adsabs.harvard.edu/abs/2005A&A...429..335V}{429, 335}

\bibitem[{{Wang} {et~al.}(2015){Wang}, {Cao}, {Liu}, {Xu}, {Liu}, {Zeng},
  {Chae}, \& {Ji}}]{Wang2015NatCo}
{Wang}, H., {Cao}, W., {Liu}, C., {et~al.} 2015,
  \href{http://dx.doi.org/10.1038/ncomms8008}{\color{magenta}Nature
  Communications},
  \href{https://ui.adsabs.harvard.edu/abs/2015NatCo...6.7008W}{6, 7008}

\bibitem[{{Wang} {et~al.}(2017){Wang}, {Liu}, {Ahn}, {Xu}, {Jing}, {Deng},
  {Huang}, {Liu}, {Kusano}, {Fleishman}, {Gary}, \& {Cao}}]{Wang2017NatAs}
{Wang}, H., {Liu}, C., {Ahn}, K., {et~al.} 2017,
  \href{http://dx.doi.org/10.1038/s41550-017-0085}{\color{magenta}Nature
  Astronomy}, \href{https://ui.adsabs.harvard.edu/abs/2017NatAs...1E..85W}{1,
  0085}

\bibitem[{{Wang} {et~al.}(2004){Wang}, {Bovik}, {Sheikh}, \&
  {Simoncelli}}]{Wang2004ITIP}
{Wang}, Z., {Bovik}, A.~C., {Sheikh}, H.~R., \& {Simoncelli}, E.~P. 2004,
  \href{http://dx.doi.org/10.1109/TIP.2003.819861}{\color{magenta}IEEE
  Transactions on Image Processing},
  \href{https://ui.adsabs.harvard.edu/abs/2004ITIP...13..600W}{13, 600}

\bibitem[{{Welsch}(2015)}]{welsch2015}
{Welsch}, B.~T. 2015,
  \href{http://dx.doi.org/10.1093/pasj/psu151}{\color{magenta}\pasj},
  \href{https://ui.adsabs.harvard.edu/abs/2015PASJ...67...18W}{67, 18}

\bibitem[{{Wyper} {et~al.}(2017){Wyper}, {Antiochos}, \&
  {DeVore}}]{Wyper2017Natur}
{Wyper}, P.~F., {Antiochos}, S.~K., \& {DeVore}, C.~R. 2017,
  \href{http://dx.doi.org/10.1038/nature22050}{\color{magenta}\nat},
  \href{https://ui.adsabs.harvard.edu/abs/2017Natur.544..452W}{544, 452}

\end{thebibliography}



\end{document}